\documentclass[twocolumn, superscriptaddress, amsfonts, amssymb, amsmath, reprint, showkeys, nofootinbib, twoside]{revtex4-2}
\usepackage[english]{babel}

\usepackage[utf8]{inputenc}
\usepackage{xcolor}
\usepackage[colorinlistoftodos, color=green!40, prependcaption]{todonotes}
\usepackage{amsthm}
\usepackage{mathtools}
\usepackage{physics}
\usepackage{xcolor}
\usepackage{graphicx}
\usepackage[left=23mm,right=13mm,top=35mm,columnsep=15pt]{geometry} 
\usepackage{adjustbox}
\usepackage{placeins}
\usepackage[T1]{fontenc}
\usepackage{lipsum}
\usepackage{csquotes}
\usepackage[pdftex, pdftitle={Article}, pdfauthor={Author}]{hyperref} 
\usepackage{caption} 
\newcounter{suppfigure}
\newenvironment{suppfigure*}{%
  \begin{figure*}[htb!]%
  \captionsetup{labelformat=simple, labelsep=period, figurename=Supplementary Figure}%
  \addtocounter{suppfigure}{1}%
}{%
  \end{figure*}%
}
\begin{document}


\title{An integrated photonics platform for high-speed, ultrahigh-extinction, many-channel quantum control}

\newcommand{\Quera}{QuEra Computing Inc, Boston, MA, 02135, USA}
\newcommand{\Sandia}{Sandia National Laboratories, Albuquerque, NM, 87185, USA}
\newcommand{\UoA}{Wyant College of Optical Sciences, University of Arizona, Tuscon, AZ, 85721, USA}
\newcommand{\MIT}{Massachusetts Institute of Technology, Cambridge, MA, 02139, USA}

\author{Mengdi Zhao}\thanks{These authors contributed equally to this work.}\affiliation{\Quera}
\author{Manuj Singh}\thanks{These authors contributed equally to this work.}\affiliation{\Quera}
\author{Anshuman Singh}\thanks{These authors contributed equally to this work.}\affiliation{\Quera}
\author{Henry Thoreen}\thanks{These authors contributed equally to this work.}\affiliation{\Quera}
\author{Robert J. DeAngelo}\affiliation{\Quera}
\author{Daniel Dominguez}\affiliation{\Sandia}
\author{Andrew Leenheer}\affiliation{\Sandia}
\author{Frédéric Peyskens} \affiliation{\Quera}
\author{Alexander Lukin} \affiliation{\Quera}
\author{Dirk Englund}\affiliation{\MIT}
\author{Matt Eichenfield}\affiliation{\Sandia}\affiliation{\UoA}
\author{Nathan Gemelke} \affiliation{\Quera}
\author{Noel H. Wan}
    \email[Corresponding author]{}
    \affiliation{\Quera}

\date{\today} 

\begin{abstract}
High-fidelity control of the thousands to millions of programmable qubits needed for utility-scale quantum computers presents a formidable challenge for control systems. In leading atomic systems, control is optical: UV–NIR beams must be fanned out over numerous spatial channels and modulated to implement gates. While photonic integrated circuits (PICs) offer a potentially scalable solution, they also need to simultaneously feature high-speed and high-extinction modulation, strong inter-channel isolation, and broad wavelength compatibility. Here, we introduce and experimentally validate a foundry-fabricated PIC platform that overcomes these limitations. Designed for Rubidium-87 neutral atom quantum computers, our 8-channel PICs, fabricated on a 200-mm wafer process, demonstrate an advanced combination of performance metrics. At the 795 nm single-qubit gate wavelength, we achieve a mean extinction ratio (ER) of 71.4 $\pm$ 1.1 dB, nearest-neighbor on-chip crosstalk of -68.0$ \pm$ 1.0 dB, and -50.8 ± 0.2 dB after parallel beam delivery in free-space. This high-performance operation extends to the 420 nm and 1013 nm wavelengths for two-qubit Rydberg gates, showing ERs of 42.4 dB (detector-limited) and 61.5 dB, respectively. The devices exhibit 10-90\% rise times of 26 $\pm$ 7\,ns, achieve dynamic switching to -60 dB levels within microsecond timescales, and show pulse stability errors at the $10^{-3}$ level. This work establishes a scalable platform for developing advanced large-scale optical control required in fault-tolerant quantum computers and other precision technologies.
\end{abstract}

\maketitle
\section{Introduction}
\begin{figure*}[ht!]
    \centering
    \includegraphics[width=1.0\textwidth]
    {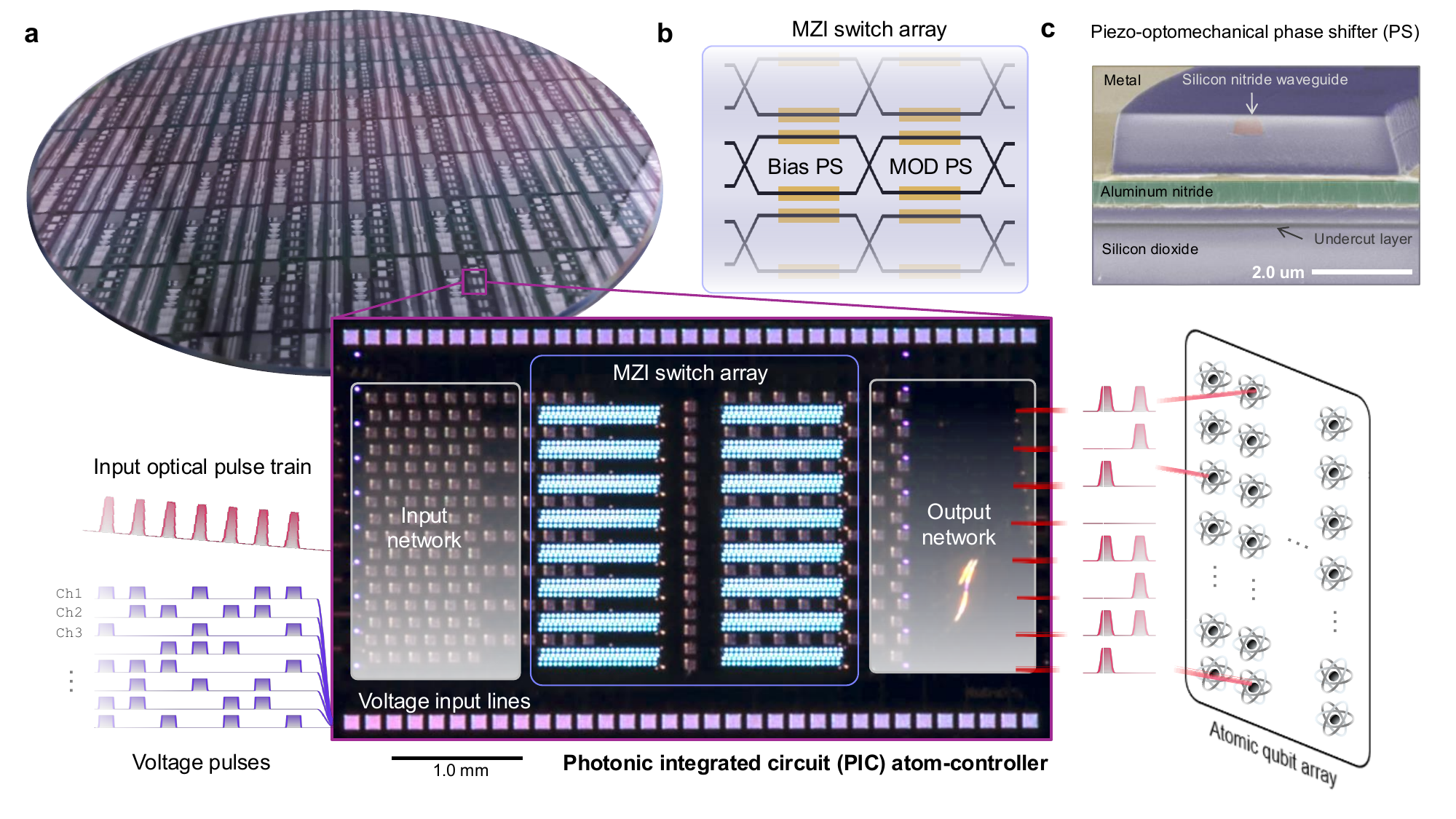}
    \caption{\textbf{A scalable integrated photonics-based architecture for high-fidelity atomic quantum control.} \textbf{a,} A fully processed 200 mm wafer for precise optical addressing of atomic qubit arrays. The central optical micrograph shows a single, multichannel atom-controller in a photonic integrated circuit (PIC) that transduces electronic waveforms into precise optical pulse trains for parallel, individual qubit addressing. \textbf{b,}As the core functional unit, each programmable MZI switch incorporates phase shifters (PS) for both static bias tuning and high-speed modulation (MOD). \textbf{c,} False color scanning electron micrograph image of the underlying piezo-optomechanical technology, which consists of an integrated aluminum nitride (AlN) actuator for high-bandwidth phase modulation in the mechanically coupled silicon nitride (SiN) waveguide.
    }
    \label{fig::concept}
\end{figure*}

\begin{figure*}[ht!]
    \centering
    \includegraphics[width=0.9\textwidth]
    {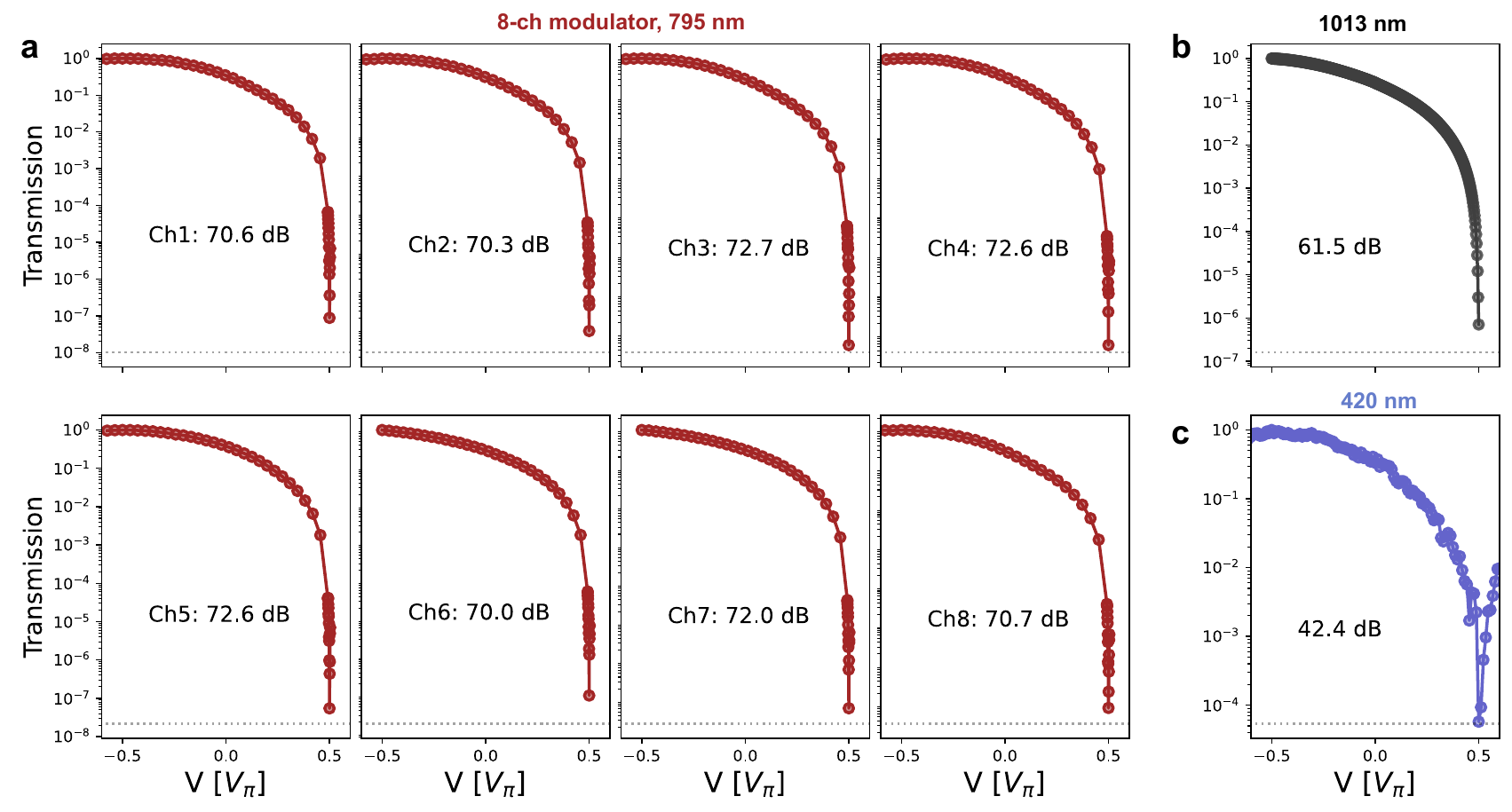}
    \caption{\textbf{Multi-channel ultrahigh extinction spanning blue to NIR}. Transmission characteristics of the Mach-Zehnder interferometer (MZI) modulators. The plots show normalized transmission versus  voltage normalized to the half-wave voltage, $V_\pi$. Dashed lines indicate measurement noise floor. \textbf{a,} Extinction ratio (ER) measurements for all 8 channels of a single PIC operating at 795 nm near the $D1$ transition of $^{87}$Rb. \textbf{b,c} Demonstration of high-contrast modulation at other key atomic transition wavelengths measured at 1013 nm and 420 nm (noise-floor limited).}
    \label{fig::ER}
\end{figure*}

\begin{figure*}[ht]
    \centering
    \includegraphics[width=1.0\textwidth]
    {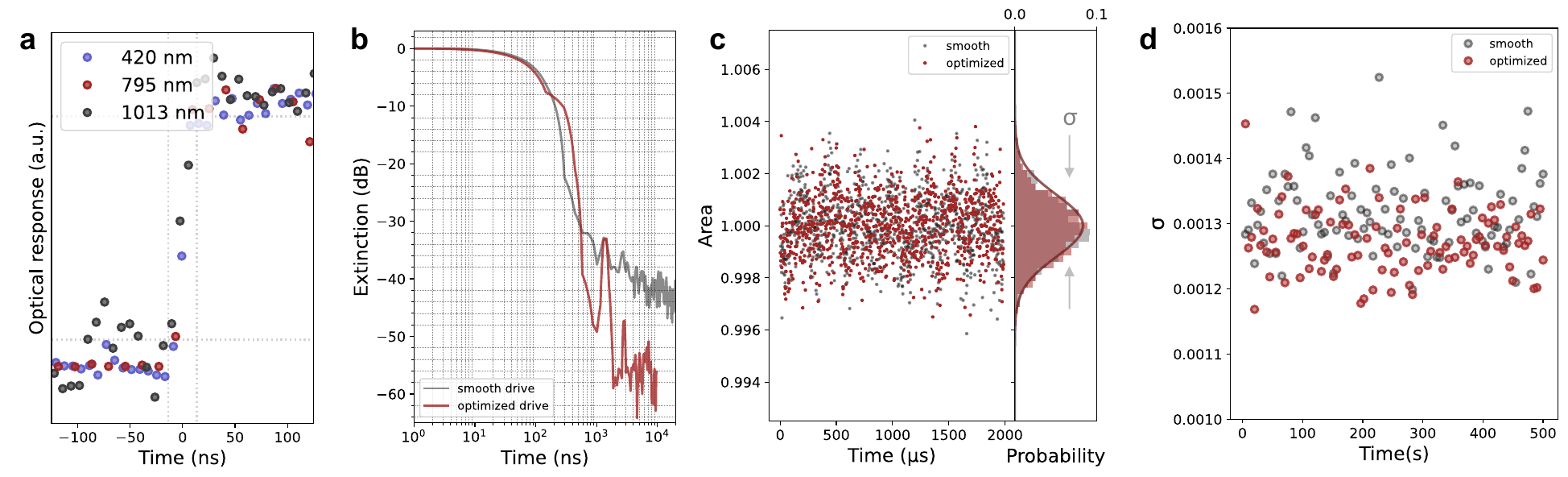}
    \caption{
    \textbf{High-speed modulation and temporal stability}. Dynamic response of the MZI modulators to high-speed electrical drive signals. \textbf{a,} Optical response at 420 nm, 795 nm, and 1013 nm, showing 10-90\% rise times of 26 $\pm$7\,ns. \textbf{b,} The dynamic extinction versus time after switching \texttt{OFF}, using a smooth drive (grey) and an optimized drive (red). \textbf{c,} Normalized pulse area of 1,000 consecutive optical pulses (left) and histogram of pulse area distribution (right). \textbf{d,} Standard deviation of pulse areas measured over 500 seconds.}
    \label{fig::temporal}
\end{figure*}

\begin{figure*}[ht]
    \centering
    \includegraphics[width=0.8\textwidth]
    {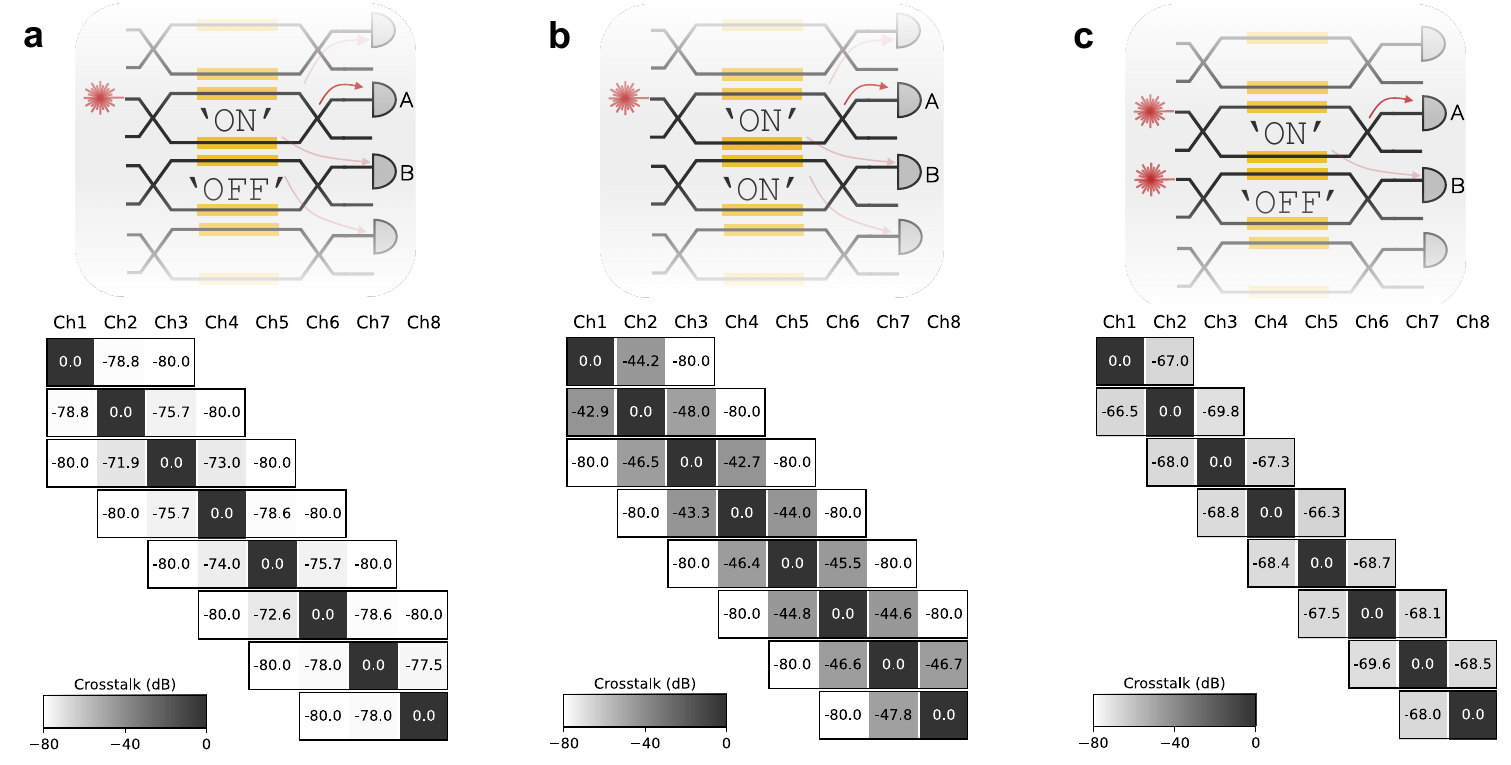}
    \caption{\textbf{On-chip optical crosstalk.} Crosstalk characterization between a primary channel and its adjacent channels under three distinct operating conditions. \textbf{a,} Device-level crosstalk of an optically-active channel in its \texttt{ON} state adjacent to an optically inactive modulator in its \texttt{OFF} state, showing nearest-neighbor (NN) and next-nearest-neighbor (NNN) crosstalk, respectively. \textbf{b,} Crosstalk matrix for an active channel in its \texttt{ON} state adjacent to an inactive modulator in its  \texttt{ON} state. \textbf{c,} Both channels are optically active, with the primary channel \texttt{ON} state and adjacent channel \texttt{OFF}.}
    \label{fig::crosstalk}
\end{figure*}

\begin{figure}[ht!]
    \centering
    \includegraphics[width=0.5\textwidth]
    {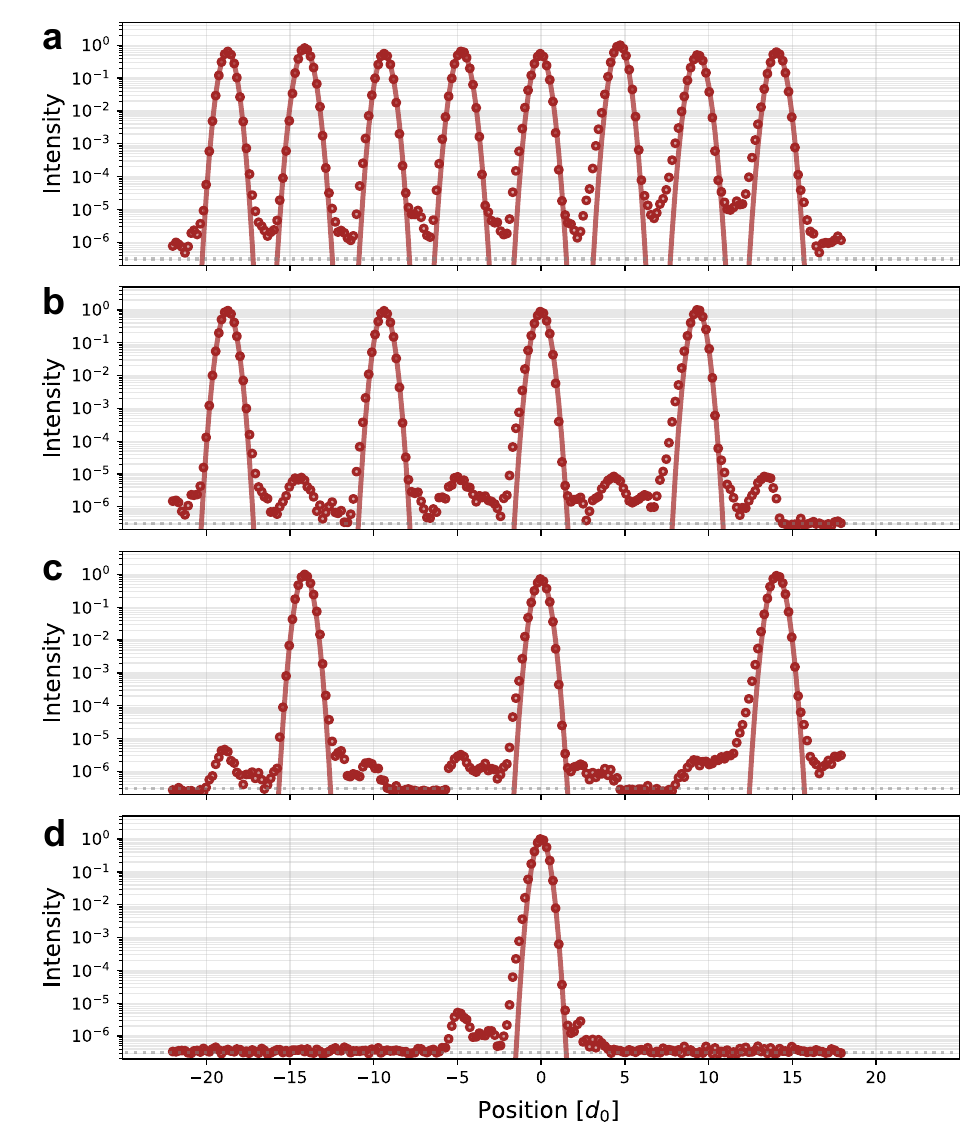}
    \caption{\textbf{System-level crosstalk in free space.} Spatial intensity profiles of the optical beams at the target plane of the free-space delivery system. All intensities are normalized to the peak, and position is normalized to the 1/$e^2$ beam diameter, $d_0$. \textbf{a,} Cross-section of the 8-channel beam array. \textbf{b-d,} Profiles for selectively addressed subsets of the array showing four \textbf{b}, two \textbf{c}, and a single \textbf{d} active channel, respectively.}
    \label{fig::beam_delivery}
\end{figure}

The development of quantum information processors capable of solving classically intractable problems is a central goal in modern science and engineering\, \cite{shor1999polynomial,nielsen2010quantum, huang2025vast}. This pursuit has driven significant advances across diverse physical platforms, where rapid improvements in system sizes and gate fidelities are now enabling demonstrations of experimental quantum error correction\,\cite{sivak2023real,ryan2024high,Bluvstein2024-rd, Bluvstein2025} and a broader range of quantum applications\,\cite{preskill2018quantum,alexeev2021quantum,dalzell2023quantum}. Among leading physical platforms, atomic systems show exceptional promise for large-scale quantum computation owing to their natural advantages in achieving large qubit numbers, dense connectivity, and massive parallelism\,\cite{saffman2010quantum,Levine2019-dz,Ebadi2021-sy,Graham2022-ne, bluvstein2022quantum, manetsch2024tweezer, Chiu2025, Bluvstein2025}. The control of these systems is fundamentally optical; arrays of precisely shaped and modulated laser pulses are required for many stages of operation, from cooling and trapping to the execution of high-fidelity quantum gates\cite{endres2016atom,ma2023high, evered2023high}. As these systems advance toward fault tolerance, the demand for controlling thousands to millions of laser modes and qubits presents a critical bottleneck\,\cite{Debnath2016-sy,zhang2024scaled}. The current paradigm, which relies on the assembly of discrete optoelectronic components, is untenable at this scale due to the prohibitive complexity and trade-offs in integration and stabilization\,\cite{Mehta2016-ak, newman2019architecture}.

Photonic integrated circuits (PICs), fabricated using scalable semiconductor techniques\,\cite{Mehta2016-ak, Niffenegger2020-up, Mehta2020-in, Wan2020-xs,hogle2023high, Isichenko2023-gr, Knollmann:24, Menon2024-lo, craft2024low,Loh2025-ci}, are a compelling solution to this control challenge. However, the requirements for quantum control are uniquely stringent,\cite{Menssen:2_optica,Palm2023-ju,Christen2025-cf}, closely motivated by the requirements for fault-tolerant quantum computation. Foremost, a viable platform must operate efficiently across a wide spectral band (e.g., UV, visible, NIR) to drive the distinct transitions for qubit control \cite{tran2022extending, Corato-Zanarella2023-bo, park2024technologies}. Switching speeds must be under the microsecond timescales to enable gate operations much faster than qubit decoherence times. Critically, the \texttt{ON/OFF}  extinction ratio (ER) of the optical modulator must be exceptionally high. Leaked photons from an \texttt{OFF} channel can drive unwanted rotations on idle qubits, leading to a gate error that scales with the leaked field or intensity\,\cite{hogle2023high}. Achieving gate infidelities below the threshold for quantum error correction (typically <$10^{-3}$\, \cite{kitaev2003fault,fowler2012surface}) would require suppressing these per-gate photonic errors to the <$10^{-4}$ level or below\,\cite{nielsen2010quantum}, necessitating an ER exceeding 50-60 dB, a benchmark that is challenging for integrated devices.

Here, we present an integrated photonic platform that satisfies these simultaneous requirements for high-fidelity quantum control. Our system, based on an integrated photonic architecture (Fig.~\ref{fig::concept}) fabricated on a 200-mm wafer process, provides multichannel optical modulation with state-of-the-art performances. We rigorously characterize these PICs at wavelengths critical for neutral-atom quantum computers based on Rubidium-87: 795 nm for single-qubit $D1$ line transitions, and 420 nm and 1013 nm for two-qubit Rydberg gate schemes. We report, to the best of our knowledge, record-high extinction ratios, as well as nanosecond switching, high dynamic contrast, low on-chip crosstalk, and high-stability pulse generation. This work establishes a robust, foundry-compatible technological foundation for the scalable control systems required by future quantum processors and other advanced photonic applications.

\section{Architecture and device design}

The control system is realized as a multichannel modulator array fabricated on a CMOS-compatible piezo-optomechanical photonics platform~\cite{stanfield2019cmos}. We utilize silicon nitride (SiN) as the waveguide core material, leveraging its wide transparency window to ensure low-loss operation from the 420 nm (blue) to the 1013 nm (NIR) wavelengths. As depicted in Fig~\ref{fig::concept}b-c, phase modulation is driven by integrated aluminum nitride (AlN) piezoelectric actuators~\cite{dong2022high, dong2022piezo} that are mechanically coupled to the SiN waveguides, enabling efficient, high-speed operation. To achieve the requisite ultrahigh extinction, each channel employs a cascaded Mach-Zehnder interferometer (MZI) architecture designed to suppress optical leakage. A single chip integrates arrays of these high-performance modulators with passive photonic networks for routing and light coupling (Fig~\ref{fig::concept}a). Supplementary Information provides additional details on wafer fabrication and chip design.

\section{Results}
A high static extinction ratio is critical for minimizing gate errors caused by parasitic light. We characterized the performance of our MZI modulators by measuring the optical transmission as a function of the applied voltage. Figure~\ref{fig::ER}a shows the normalized transmission for all 8 channels of a single PIC operating at 795 nm. Every channel exhibits an ER exceeding 70 dB, with a mean value of 71.4 $\pm$ 1.1 dB, which surpasses reported values for integrated photonic modulators. This high-contrast performance is achieved with a single-ended half-wave voltage $V_{\pi}$ of 74.7 $\pm$ 3.7\,V. We demonstrate the platform's crucial multi-wavelength capability by characterizing devices designed for two-qubit gate wavelengths. At 1013 nm, we measure an ER of 61.5 dB (Fig.~\ref{fig::ER}b, $V_{\pi} = 200$ V), and at 420 nm, we achieve an ER of 42.4 dB (Fig.~\ref{fig::ER}c, $V_{\pi} = 44.4$ V), a value limited by the noise floor of our detection setup.

High-fidelity quantum operations demand not only high static contrast but also fast, stable, and high-contrast dynamic switching. We first characterized the intrinsic speed of the piezo-optomechanical actuators. As shown in Fig~\ref{fig::temporal}a, the optical response exhibits a 10-90\% rise time of 26\,$\pm$\,7\,ns for devices measured at 420 nm, 795 nm and 1013 nm. Achieving high dynamic extinction during switching requires careful control of the drive signal. We found that a simple drive pulse limits the dynamic ER, whereas an optimized drive waveform, pre-distorted to compensate for the system's impulse response, suppresses optical transients and achieves a dynamic extinction below $10^{-6}$ within one microsecond of switching, as shown in Fig\,\ref{fig::temporal}b. The stability of the resulting optical pulses is critical for consistent gate operations. An analysis of 1,000 consecutive pulses shows a normalized pulse area standard deviation of 0.1\% (Fig.~\ref{fig::temporal}c). This stability is maintained over extended periods, with a mean sample standard deviation of 0.13\% measured over 500 seconds of continuous operation (Fig.~\ref{fig::temporal}d).

High optical isolation between channels is required to prevent addressing errors in scalable architectures. We quantified the on-chip optical crosstalk by measuring the power leakage between adjacent modulators under various operating conditions. Figure~\ref{fig::crosstalk}a shows the crosstalk matrix from an optically active \texttt{ON} channel into an adjacent channel with no optical input and its modulator set to the \texttt{OFF} state. This configuration considers all possible mechanisms of leakage; the nearest-neighbor (NN) crosstalk is -76.2 $\pm$ 2.4 dB, and the next-nearest-neighbor (NNN) crosstalk is below the -80 dB measurement floor. Figure Figure~\ref{fig::crosstalk}b shows a worst-case scenario where the adjacent modulator is in the \texttt{ON} state, resulting in a higher NN crosstalk of -45.3 $\pm$ 1.7. Another operationally relevant configuration, where both the primary and adjacent channels are optically active and switched to \texttt{ON} and \texttt{OFF} states, respectively, we measured a mean nearest-neighbor (NN) crosstalk of -68.0 $\pm$ 1.0 dB (Fig.~\ref{fig::crosstalk}).

In addition to on-chip leakage, overall system crosstalk is determined by the free-space propagation of the output beams projected onto the qubit array. We characterized this system-level crosstalk by imaging the output of the PIC through a beam delivery system designed to produce an array of focused spots. Figure~\ref{fig::beam_delivery}a shows well-formed Gaussian spatial intensity profile of the 8-channel beam array at the target plane, with a pitch of 4.33 $d_0$ (where $d_0$ is the 1/$e^2$ beam diameter). By selectively activating subsets of channels, we find that the dominant contribution to leakage at an NN idle qubit location arises from evanescent coupling at -50.8 ± 0.16 dB, as shown in Fig.\,\ref{fig::beam_delivery}b, while contributions to NNN channels are below the measurement noise floor of -65 dB.

Finally, long-term operational stability and high power handling are essential for practical deployment. At 795 nm, the devices exhibited robust performance, showing no signs of degradation or instability while injecting over 1\,W of total input power (approximately 100\,mW per channel) over months of continuous experiments. To counteract slow thermal drifts, we implemented a 5 Hz feedback loop that actively stabilizes the modulator bias point. This system successfully locked the extinction to 69.8$\pm$3.0 dB over a 20-hour period (see Supplementary Information). At 420\,nm, we observed the onset of slow power fluctuations at input powers above 10 mW, suggesting a power-dependent instability. Investigation into the origin and mitigation of this short-wavelength instability is a direction for future work.

\section{Discussion}
The results presented here constitute a significant advance in integrated optical and quantum control technology, establishing a platform that meets the requirements for high-fidelity quantum computing. The demonstrated suite of advanced performance metrics provides a direct path to scaling the number of independent optical channels in leading quantum computing hardware architectures. Furthermore, the deployment of these PIC modules into quantum computing testbeds will enable benchmarking of gate fidelities and the execution quantum algorithms based on these expanded degrees of freedom. 

Building upon this work, several key avenues for optimization can further enhance performance and scalability. Probing the intrinsic device limits for blue-UV operation, currently constrained by measurement noise floor, requires characterization of lower-loss devices. In the chips reported in this work, we measured insertion losses of 3 dB per modulator, single-mode propagation losses of 1.5 dB cm$^{-1}$, 2.7 dB cm$^{-1}$ and 5.6 dB cm$^{-1}$ at 795 nm, 1013\,nm and 420\,nm, and input-output chip coupling losses of ranging between -3 to -10 dB. While atomic quantum computing does not require ultra-low-loss photonics, reducing the total losses will decrease laser power overhead and effectively enable larger channel counts. Future work will target these loss mechanisms through improved device designs and coupling schemes, with new material integration also offering another promising path for optimization at short wavelengths~\cite{west2019low, castillo2024cmos}. Finally, future design iterations will explore differential drive configurations and more efficient phase shifter geometries to lower the required voltage, thereby simplifying the driver requirements towards dense integration\,\cite{atabaki2018integrating, baehr2023monolithically}

In conclusion, we have developed and rigorously characterized a photonic control system with multi-channel operation at key atomic transition wavelengths. By achieving high-speed switching, record extinction ratios, excellent channel isolation, and high pulse stability, we have established an integrated photonic technology for manipulating quantum systems at scale. This work provides a foundational architecture for the complex optical control systems required to build fault-tolerant quantum computers and opens new possibilities for advanced applications in precision metrology, optical displays, and biophotonics.

\section{Acknowledgments}
This work is supported by the Defense Advanced Research Projects Agency (DARPA) through the ONISQ program (W911NF2010021), STTR program (140D0422C0035) and OASIC program (HR0011-24-P-0314), the Air Force Research Laboratory (AFOSR) STTR program (FA8750-20-P-1706) and the BIRD Foundation (Project \#1833). Sandia National Laboratories is a multi-mission laboratory managed and operated by National Technology \& Engineering Solutions of Sandia, LLC (NTESS), a wholly owned subsidiary of Honeywell International Inc., for the U.S. Department of Energy’s National Nuclear Security Administration (DOE/NNSA) under contract DE-NA0003525. This written work is authored by an employee of NTESS. The employee, not NTESS, owns the right, title and interest in and to the written work and is responsible for its contents. Any subjective views or opinions that might be expressed in the written work do not necessarily represent the views of the U.S. Government. The publisher acknowledges that the U.S. Government retains a non-exclusive, paid-up, irrevocable, world-wide license to publish or reproduce the published form of this written work or allow others to do so, for U.S. Government purposes. The DOE will provide public access to results of federally sponsored research in accordance with the DOE Public Access Plan. We acknowledge support from the QuEra Computing team especially contributions from Ming-Guang Hu, Donggyu Kim, Spencer Haney, Minho Kwon and Tom Karolyshlyn. 

\bibliography{biblio}

\begin{thebibliography}{48}%
\makeatletter
\providecommand \@ifxundefined [1]{%
 \@ifx{#1\undefined}
}%
\providecommand \@ifnum [1]{%
 \ifnum #1\expandafter \@firstoftwo
 \else \expandafter \@secondoftwo
 \fi
}%
\providecommand \@ifx [1]{%
 \ifx #1\expandafter \@firstoftwo
 \else \expandafter \@secondoftwo
 \fi
}%
\providecommand \natexlab [1]{#1}%
\providecommand \enquote  [1]{``#1''}%
\providecommand \bibnamefont  [1]{#1}%
\providecommand \bibfnamefont [1]{#1}%
\providecommand \citenamefont [1]{#1}%
\providecommand \href@noop [0]{\@secondoftwo}%
\providecommand \href [0]{\begingroup \@sanitize@url \@href}%
\providecommand \@href[1]{\@@startlink{#1}\@@href}%
\providecommand \@@href[1]{\endgroup#1\@@endlink}%
\providecommand \@sanitize@url [0]{\catcode `\\12\catcode `\$12\catcode `\&12\catcode `\#12\catcode `\^12\catcode `\_12\catcode `\%12\relax}%
\providecommand \@@startlink[1]{}%
\providecommand \@@endlink[0]{}%
\providecommand \url  [0]{\begingroup\@sanitize@url \@url }%
\providecommand \@url [1]{\endgroup\@href {#1}{\urlprefix }}%
\providecommand \urlprefix  [0]{URL }%
\providecommand \Eprint [0]{\href }%
\providecommand \doibase [0]{http://dx.doi.org/}%
\providecommand \selectlanguage [0]{\@gobble}%
\providecommand \bibinfo  [0]{\@secondoftwo}%
\providecommand \bibfield  [0]{\@secondoftwo}%
\providecommand \translation [1]{[#1]}%
\providecommand \BibitemOpen [0]{}%
\providecommand \bibitemStop [0]{}%
\providecommand \bibitemNoStop [0]{.\EOS\space}%
\providecommand \EOS [0]{\spacefactor3000\relax}%
\providecommand \BibitemShut  [1]{\csname bibitem#1\endcsname}%
\let\auto@bib@innerbib\@empty
\bibitem [{\citenamefont {Shor}(1999)}]{shor1999polynomial}%
  \BibitemOpen
  \bibfield  {author} {\bibinfo {author} {\bibfnamefont {P.~W.}\ \bibnamefont {Shor}},\ }\href@noop {} {\bibfield  {journal} {\bibinfo  {journal} {SIAM review}\ }\textbf {\bibinfo {volume} {41}},\ \bibinfo {pages} {303} (\bibinfo {year} {1999})}\BibitemShut {NoStop}%
\bibitem [{\citenamefont {Nielsen}\ and\ \citenamefont {Chuang}(2010)}]{nielsen2010quantum}%
  \BibitemOpen
  \bibfield  {author} {\bibinfo {author} {\bibfnamefont {M.~A.}\ \bibnamefont {Nielsen}}\ and\ \bibinfo {author} {\bibfnamefont {I.~L.}\ \bibnamefont {Chuang}},\ }\href@noop {} {\emph {\bibinfo {title} {Quantum computation and quantum information}}}\ (\bibinfo  {publisher} {Cambridge university press},\ \bibinfo {year} {2010})\BibitemShut {NoStop}%
\bibitem [{\citenamefont {Huang}\ \emph {et~al.}(2025)\citenamefont {Huang}, \citenamefont {Choi}, \citenamefont {McClean},\ and\ \citenamefont {Preskill}}]{huang2025vast}%
  \BibitemOpen
  \bibfield  {author} {\bibinfo {author} {\bibfnamefont {H.-Y.}\ \bibnamefont {Huang}}, \bibinfo {author} {\bibfnamefont {S.}~\bibnamefont {Choi}}, \bibinfo {author} {\bibfnamefont {J.~R.}\ \bibnamefont {McClean}}, \ and\ \bibinfo {author} {\bibfnamefont {J.}~\bibnamefont {Preskill}},\ }\href {\doibase 10.48550/arXiv.2508.05720} {\enquote {\bibinfo {title} {The vast world of quantum advantage},}\ } (\bibinfo {year} {2025}),\ \Eprint {http://arxiv.org/abs/2508.05720} {arXiv:2508.05720 [quant-ph]} \BibitemShut {NoStop}%
\bibitem [{\citenamefont {Sivak}\ \emph {et~al.}(2023)\citenamefont {Sivak}, \citenamefont {Eickbusch}, \citenamefont {Royer}, \citenamefont {Singh}, \citenamefont {Tsioutsios}, \citenamefont {Ganjam}, \citenamefont {Miano}, \citenamefont {Brock}, \citenamefont {Ding}, \citenamefont {Frunzio} \emph {et~al.}}]{sivak2023real}%
  \BibitemOpen
  \bibfield  {author} {\bibinfo {author} {\bibfnamefont {V.~V.}\ \bibnamefont {Sivak}}, \bibinfo {author} {\bibfnamefont {A.}~\bibnamefont {Eickbusch}}, \bibinfo {author} {\bibfnamefont {B.}~\bibnamefont {Royer}}, \bibinfo {author} {\bibfnamefont {S.}~\bibnamefont {Singh}}, \bibinfo {author} {\bibfnamefont {I.}~\bibnamefont {Tsioutsios}}, \bibinfo {author} {\bibfnamefont {S.}~\bibnamefont {Ganjam}}, \bibinfo {author} {\bibfnamefont {A.}~\bibnamefont {Miano}}, \bibinfo {author} {\bibfnamefont {B.}~\bibnamefont {Brock}}, \bibinfo {author} {\bibfnamefont {A.}~\bibnamefont {Ding}}, \bibinfo {author} {\bibfnamefont {L.}~\bibnamefont {Frunzio}},  \emph {et~al.},\ }\href@noop {} {\bibfield  {journal} {\bibinfo  {journal} {Nature}\ }\textbf {\bibinfo {volume} {616}},\ \bibinfo {pages} {50} (\bibinfo {year} {2023})}\BibitemShut {NoStop}%
\bibitem [{\citenamefont {Ryan-Anderson}\ \emph {et~al.}(2024)\citenamefont {Ryan-Anderson}, \citenamefont {Brown}, \citenamefont {Baldwin}, \citenamefont {Dreiling}, \citenamefont {Foltz}, \citenamefont {Gaebler}, \citenamefont {Gatterman}, \citenamefont {Hewitt}, \citenamefont {Holliman}, \citenamefont {Horst} \emph {et~al.}}]{ryan2024high}%
  \BibitemOpen
  \bibfield  {author} {\bibinfo {author} {\bibfnamefont {C.}~\bibnamefont {Ryan-Anderson}}, \bibinfo {author} {\bibfnamefont {N.}~\bibnamefont {Brown}}, \bibinfo {author} {\bibfnamefont {C.}~\bibnamefont {Baldwin}}, \bibinfo {author} {\bibfnamefont {J.}~\bibnamefont {Dreiling}}, \bibinfo {author} {\bibfnamefont {C.}~\bibnamefont {Foltz}}, \bibinfo {author} {\bibfnamefont {J.}~\bibnamefont {Gaebler}}, \bibinfo {author} {\bibfnamefont {T.}~\bibnamefont {Gatterman}}, \bibinfo {author} {\bibfnamefont {N.}~\bibnamefont {Hewitt}}, \bibinfo {author} {\bibfnamefont {C.}~\bibnamefont {Holliman}}, \bibinfo {author} {\bibfnamefont {C.}~\bibnamefont {Horst}},  \emph {et~al.},\ }\href@noop {} {\bibfield  {journal} {\bibinfo  {journal} {Science}\ }\textbf {\bibinfo {volume} {385}},\ \bibinfo {pages} {1327} (\bibinfo {year} {2024})}\BibitemShut {NoStop}%
\bibitem [{\citenamefont {Bluvstein}\ \emph {et~al.}(2024)\citenamefont {Bluvstein}, \citenamefont {Evered}, \citenamefont {Geim}, \citenamefont {Li}, \citenamefont {Zhou}, \citenamefont {Manovitz}, \citenamefont {Ebadi}, \citenamefont {Cain}, \citenamefont {Kalinowski}, \citenamefont {Hangleiter}, \citenamefont {Bonilla~Ataides}, \citenamefont {Maskara}, \citenamefont {Cong}, \citenamefont {Gao}, \citenamefont {Sales~Rodriguez}, \citenamefont {Karolyshyn}, \citenamefont {Semeghini}, \citenamefont {Gullans}, \citenamefont {Greiner}, \citenamefont {Vuleti{\'c}},\ and\ \citenamefont {Lukin}}]{Bluvstein2024-rd}%
  \BibitemOpen
  \bibfield  {author} {\bibinfo {author} {\bibfnamefont {D.}~\bibnamefont {Bluvstein}}, \bibinfo {author} {\bibfnamefont {S.~J.}\ \bibnamefont {Evered}}, \bibinfo {author} {\bibfnamefont {A.~A.}\ \bibnamefont {Geim}}, \bibinfo {author} {\bibfnamefont {S.~H.}\ \bibnamefont {Li}}, \bibinfo {author} {\bibfnamefont {H.}~\bibnamefont {Zhou}}, \bibinfo {author} {\bibfnamefont {T.}~\bibnamefont {Manovitz}}, \bibinfo {author} {\bibfnamefont {S.}~\bibnamefont {Ebadi}}, \bibinfo {author} {\bibfnamefont {M.}~\bibnamefont {Cain}}, \bibinfo {author} {\bibfnamefont {M.}~\bibnamefont {Kalinowski}}, \bibinfo {author} {\bibfnamefont {D.}~\bibnamefont {Hangleiter}}, \bibinfo {author} {\bibfnamefont {J.~P.}\ \bibnamefont {Bonilla~Ataides}}, \bibinfo {author} {\bibfnamefont {N.}~\bibnamefont {Maskara}}, \bibinfo {author} {\bibfnamefont {I.}~\bibnamefont {Cong}}, \bibinfo {author} {\bibfnamefont {X.}~\bibnamefont {Gao}}, \bibinfo {author} {\bibfnamefont {P.}~\bibnamefont {Sales~Rodriguez}}, \bibinfo {author} {\bibfnamefont
  {T.}~\bibnamefont {Karolyshyn}}, \bibinfo {author} {\bibfnamefont {G.}~\bibnamefont {Semeghini}}, \bibinfo {author} {\bibfnamefont {M.~J.}\ \bibnamefont {Gullans}}, \bibinfo {author} {\bibfnamefont {M.}~\bibnamefont {Greiner}}, \bibinfo {author} {\bibfnamefont {V.}~\bibnamefont {Vuleti{\'c}}}, \ and\ \bibinfo {author} {\bibfnamefont {M.~D.}\ \bibnamefont {Lukin}},\ }\href@noop {} {\bibfield  {journal} {\bibinfo  {journal} {Nature}\ }\textbf {\bibinfo {volume} {626}},\ \bibinfo {pages} {58} (\bibinfo {year} {2024})}\BibitemShut {NoStop}%
\bibitem [{\citenamefont {Bluvstein}\ \emph {et~al.}(2025)\citenamefont {Bluvstein}, \citenamefont {Geim} \emph {et~al.}}]{Bluvstein2025}%
  \BibitemOpen
  \bibfield  {author} {\bibinfo {author} {\bibfnamefont {D.}~\bibnamefont {Bluvstein}}, \bibinfo {author} {\bibfnamefont {A.~A.}\ \bibnamefont {Geim}},  \emph {et~al.},\ }\href {https://arxiv.org/abs/2506.20661} {\bibfield  {journal} {\bibinfo  {journal} {arXiv preprint arXiv:2506.20661}\ } (\bibinfo {year} {2025})},\ \Eprint {http://arxiv.org/abs/2506.20661} {arXiv:2506.20661 [quant-ph]} \BibitemShut {NoStop}%
\bibitem [{\citenamefont {Preskill}(2018)}]{preskill2018quantum}%
  \BibitemOpen
  \bibfield  {author} {\bibinfo {author} {\bibfnamefont {J.}~\bibnamefont {Preskill}},\ }\href@noop {} {\bibfield  {journal} {\bibinfo  {journal} {Quantum}\ }\textbf {\bibinfo {volume} {2}},\ \bibinfo {pages} {79} (\bibinfo {year} {2018})}\BibitemShut {NoStop}%
\bibitem [{\citenamefont {Alexeev}\ \emph {et~al.}(2021)\citenamefont {Alexeev}, \citenamefont {Bacon}, \citenamefont {Brown}, \citenamefont {Calderbank}, \citenamefont {Carr}, \citenamefont {Chong}, \citenamefont {DeMarco}, \citenamefont {Englund}, \citenamefont {Farhi}, \citenamefont {Fefferman} \emph {et~al.}}]{alexeev2021quantum}%
  \BibitemOpen
  \bibfield  {author} {\bibinfo {author} {\bibfnamefont {Y.}~\bibnamefont {Alexeev}}, \bibinfo {author} {\bibfnamefont {D.}~\bibnamefont {Bacon}}, \bibinfo {author} {\bibfnamefont {K.~R.}\ \bibnamefont {Brown}}, \bibinfo {author} {\bibfnamefont {R.}~\bibnamefont {Calderbank}}, \bibinfo {author} {\bibfnamefont {L.~D.}\ \bibnamefont {Carr}}, \bibinfo {author} {\bibfnamefont {F.~T.}\ \bibnamefont {Chong}}, \bibinfo {author} {\bibfnamefont {B.}~\bibnamefont {DeMarco}}, \bibinfo {author} {\bibfnamefont {D.}~\bibnamefont {Englund}}, \bibinfo {author} {\bibfnamefont {E.}~\bibnamefont {Farhi}}, \bibinfo {author} {\bibfnamefont {B.}~\bibnamefont {Fefferman}},  \emph {et~al.},\ }\href@noop {} {\bibfield  {journal} {\bibinfo  {journal} {PRX quantum}\ }\textbf {\bibinfo {volume} {2}},\ \bibinfo {pages} {017001} (\bibinfo {year} {2021})}\BibitemShut {NoStop}%
\bibitem [{\citenamefont {Dalzell}\ \emph {et~al.}(2023)\citenamefont {Dalzell}, \citenamefont {McArdle}, \citenamefont {Berta}, \citenamefont {Bienias}, \citenamefont {Chen}, \citenamefont {Gily{\'e}n}, \citenamefont {Hann}, \citenamefont {Kastoryano}, \citenamefont {Khabiboulline}, \citenamefont {Kubica} \emph {et~al.}}]{dalzell2023quantum}%
  \BibitemOpen
  \bibfield  {author} {\bibinfo {author} {\bibfnamefont {A.~M.}\ \bibnamefont {Dalzell}}, \bibinfo {author} {\bibfnamefont {S.}~\bibnamefont {McArdle}}, \bibinfo {author} {\bibfnamefont {M.}~\bibnamefont {Berta}}, \bibinfo {author} {\bibfnamefont {P.}~\bibnamefont {Bienias}}, \bibinfo {author} {\bibfnamefont {C.-F.}\ \bibnamefont {Chen}}, \bibinfo {author} {\bibfnamefont {A.}~\bibnamefont {Gily{\'e}n}}, \bibinfo {author} {\bibfnamefont {C.~T.}\ \bibnamefont {Hann}}, \bibinfo {author} {\bibfnamefont {M.~J.}\ \bibnamefont {Kastoryano}}, \bibinfo {author} {\bibfnamefont {E.~T.}\ \bibnamefont {Khabiboulline}}, \bibinfo {author} {\bibfnamefont {A.}~\bibnamefont {Kubica}},  \emph {et~al.},\ }\href@noop {} {\bibfield  {journal} {\bibinfo  {journal} {arXiv preprint arXiv:2310.03011}\ } (\bibinfo {year} {2023})}\BibitemShut {NoStop}%
\bibitem [{\citenamefont {Saffman}\ \emph {et~al.}(2010)\citenamefont {Saffman}, \citenamefont {Walker},\ and\ \citenamefont {M{\o}lmer}}]{saffman2010quantum}%
  \BibitemOpen
  \bibfield  {author} {\bibinfo {author} {\bibfnamefont {M.}~\bibnamefont {Saffman}}, \bibinfo {author} {\bibfnamefont {T.~G.}\ \bibnamefont {Walker}}, \ and\ \bibinfo {author} {\bibfnamefont {K.}~\bibnamefont {M{\o}lmer}},\ }\href@noop {} {\bibfield  {journal} {\bibinfo  {journal} {Reviews of modern physics}\ }\textbf {\bibinfo {volume} {82}},\ \bibinfo {pages} {2313} (\bibinfo {year} {2010})}\BibitemShut {NoStop}%
\bibitem [{\citenamefont {Levine}\ \emph {et~al.}(2019)\citenamefont {Levine}, \citenamefont {Keesling}, \citenamefont {Semeghini}, \citenamefont {Omran}, \citenamefont {Wang}, \citenamefont {Ebadi}, \citenamefont {Bernien}, \citenamefont {Greiner}, \citenamefont {Vuleti{\'c}}, \citenamefont {Pichler},\ and\ \citenamefont {Lukin}}]{Levine2019-dz}%
  \BibitemOpen
  \bibfield  {author} {\bibinfo {author} {\bibfnamefont {H.}~\bibnamefont {Levine}}, \bibinfo {author} {\bibfnamefont {A.}~\bibnamefont {Keesling}}, \bibinfo {author} {\bibfnamefont {G.}~\bibnamefont {Semeghini}}, \bibinfo {author} {\bibfnamefont {A.}~\bibnamefont {Omran}}, \bibinfo {author} {\bibfnamefont {T.~T.}\ \bibnamefont {Wang}}, \bibinfo {author} {\bibfnamefont {S.}~\bibnamefont {Ebadi}}, \bibinfo {author} {\bibfnamefont {H.}~\bibnamefont {Bernien}}, \bibinfo {author} {\bibfnamefont {M.}~\bibnamefont {Greiner}}, \bibinfo {author} {\bibfnamefont {V.}~\bibnamefont {Vuleti{\'c}}}, \bibinfo {author} {\bibfnamefont {H.}~\bibnamefont {Pichler}}, \ and\ \bibinfo {author} {\bibfnamefont {M.~D.}\ \bibnamefont {Lukin}},\ }\href@noop {} {\bibfield  {journal} {\bibinfo  {journal} {Phys. Rev. Lett.}\ }\textbf {\bibinfo {volume} {123}},\ \bibinfo {pages} {170503} (\bibinfo {year} {2019})}\BibitemShut {NoStop}%
\bibitem [{\citenamefont {Ebadi}\ \emph {et~al.}(2021)\citenamefont {Ebadi}, \citenamefont {Wang}, \citenamefont {Levine}, \citenamefont {Keesling}, \citenamefont {Semeghini}, \citenamefont {Omran}, \citenamefont {Bluvstein}, \citenamefont {Samajdar}, \citenamefont {Pichler}, \citenamefont {Ho}, \citenamefont {Choi}, \citenamefont {Sachdev}, \citenamefont {Greiner}, \citenamefont {Vuleti{\'c}},\ and\ \citenamefont {Lukin}}]{Ebadi2021-sy}%
  \BibitemOpen
  \bibfield  {author} {\bibinfo {author} {\bibfnamefont {S.}~\bibnamefont {Ebadi}}, \bibinfo {author} {\bibfnamefont {T.~T.}\ \bibnamefont {Wang}}, \bibinfo {author} {\bibfnamefont {H.}~\bibnamefont {Levine}}, \bibinfo {author} {\bibfnamefont {A.}~\bibnamefont {Keesling}}, \bibinfo {author} {\bibfnamefont {G.}~\bibnamefont {Semeghini}}, \bibinfo {author} {\bibfnamefont {A.}~\bibnamefont {Omran}}, \bibinfo {author} {\bibfnamefont {D.}~\bibnamefont {Bluvstein}}, \bibinfo {author} {\bibfnamefont {R.}~\bibnamefont {Samajdar}}, \bibinfo {author} {\bibfnamefont {H.}~\bibnamefont {Pichler}}, \bibinfo {author} {\bibfnamefont {W.~W.}\ \bibnamefont {Ho}}, \bibinfo {author} {\bibfnamefont {S.}~\bibnamefont {Choi}}, \bibinfo {author} {\bibfnamefont {S.}~\bibnamefont {Sachdev}}, \bibinfo {author} {\bibfnamefont {M.}~\bibnamefont {Greiner}}, \bibinfo {author} {\bibfnamefont {V.}~\bibnamefont {Vuleti{\'c}}}, \ and\ \bibinfo {author} {\bibfnamefont {M.~D.}\ \bibnamefont {Lukin}},\ }\href@noop {} {\bibfield  {journal}
  {\bibinfo  {journal} {Nature}\ }\textbf {\bibinfo {volume} {595}},\ \bibinfo {pages} {227} (\bibinfo {year} {2021})}\BibitemShut {NoStop}%
\bibitem [{\citenamefont {Graham}\ \emph {et~al.}(2022)\citenamefont {Graham}, \citenamefont {Song}, \citenamefont {Scott}, \citenamefont {Poole}, \citenamefont {Phuttitarn}, \citenamefont {Jooya}, \citenamefont {Eichler}, \citenamefont {Jiang}, \citenamefont {Marra}, \citenamefont {Grinkemeyer}, \citenamefont {Kwon}, \citenamefont {Ebert}, \citenamefont {Cherek}, \citenamefont {Lichtman}, \citenamefont {Gillette}, \citenamefont {Gilbert}, \citenamefont {Bowman}, \citenamefont {Ballance}, \citenamefont {Campbell}, \citenamefont {Dahl}, \citenamefont {Crawford}, \citenamefont {Blunt}, \citenamefont {Rogers}, \citenamefont {Noel},\ and\ \citenamefont {Saffman}}]{Graham2022-ne}%
  \BibitemOpen
  \bibfield  {author} {\bibinfo {author} {\bibfnamefont {T.~M.}\ \bibnamefont {Graham}}, \bibinfo {author} {\bibfnamefont {Y.}~\bibnamefont {Song}}, \bibinfo {author} {\bibfnamefont {J.}~\bibnamefont {Scott}}, \bibinfo {author} {\bibfnamefont {C.}~\bibnamefont {Poole}}, \bibinfo {author} {\bibfnamefont {L.}~\bibnamefont {Phuttitarn}}, \bibinfo {author} {\bibfnamefont {K.}~\bibnamefont {Jooya}}, \bibinfo {author} {\bibfnamefont {P.}~\bibnamefont {Eichler}}, \bibinfo {author} {\bibfnamefont {X.}~\bibnamefont {Jiang}}, \bibinfo {author} {\bibfnamefont {A.}~\bibnamefont {Marra}}, \bibinfo {author} {\bibfnamefont {B.}~\bibnamefont {Grinkemeyer}}, \bibinfo {author} {\bibfnamefont {M.}~\bibnamefont {Kwon}}, \bibinfo {author} {\bibfnamefont {M.}~\bibnamefont {Ebert}}, \bibinfo {author} {\bibfnamefont {J.}~\bibnamefont {Cherek}}, \bibinfo {author} {\bibfnamefont {M.~T.}\ \bibnamefont {Lichtman}}, \bibinfo {author} {\bibfnamefont {M.}~\bibnamefont {Gillette}}, \bibinfo {author} {\bibfnamefont {J.}~\bibnamefont
  {Gilbert}}, \bibinfo {author} {\bibfnamefont {D.}~\bibnamefont {Bowman}}, \bibinfo {author} {\bibfnamefont {T.}~\bibnamefont {Ballance}}, \bibinfo {author} {\bibfnamefont {C.}~\bibnamefont {Campbell}}, \bibinfo {author} {\bibfnamefont {E.~D.}\ \bibnamefont {Dahl}}, \bibinfo {author} {\bibfnamefont {O.}~\bibnamefont {Crawford}}, \bibinfo {author} {\bibfnamefont {N.~S.}\ \bibnamefont {Blunt}}, \bibinfo {author} {\bibfnamefont {B.}~\bibnamefont {Rogers}}, \bibinfo {author} {\bibfnamefont {T.}~\bibnamefont {Noel}}, \ and\ \bibinfo {author} {\bibfnamefont {M.}~\bibnamefont {Saffman}},\ }\href@noop {} {\bibfield  {journal} {\bibinfo  {journal} {Nature}\ }\textbf {\bibinfo {volume} {604}},\ \bibinfo {pages} {457} (\bibinfo {year} {2022})}\BibitemShut {NoStop}%
\bibitem [{\citenamefont {Bluvstein}\ \emph {et~al.}(2022)\citenamefont {Bluvstein}, \citenamefont {Levine}, \citenamefont {Semeghini}, \citenamefont {Wang}, \citenamefont {Ebadi}, \citenamefont {Kalinowski}, \citenamefont {Keesling}, \citenamefont {Maskara}, \citenamefont {Pichler}, \citenamefont {Greiner} \emph {et~al.}}]{bluvstein2022quantum}%
  \BibitemOpen
  \bibfield  {author} {\bibinfo {author} {\bibfnamefont {D.}~\bibnamefont {Bluvstein}}, \bibinfo {author} {\bibfnamefont {H.}~\bibnamefont {Levine}}, \bibinfo {author} {\bibfnamefont {G.}~\bibnamefont {Semeghini}}, \bibinfo {author} {\bibfnamefont {T.~T.}\ \bibnamefont {Wang}}, \bibinfo {author} {\bibfnamefont {S.}~\bibnamefont {Ebadi}}, \bibinfo {author} {\bibfnamefont {M.}~\bibnamefont {Kalinowski}}, \bibinfo {author} {\bibfnamefont {A.}~\bibnamefont {Keesling}}, \bibinfo {author} {\bibfnamefont {N.}~\bibnamefont {Maskara}}, \bibinfo {author} {\bibfnamefont {H.}~\bibnamefont {Pichler}}, \bibinfo {author} {\bibfnamefont {M.}~\bibnamefont {Greiner}},  \emph {et~al.},\ }\href@noop {} {\bibfield  {journal} {\bibinfo  {journal} {Nature}\ }\textbf {\bibinfo {volume} {604}},\ \bibinfo {pages} {451} (\bibinfo {year} {2022})}\BibitemShut {NoStop}%
\bibitem [{\citenamefont {Manetsch}\ \emph {et~al.}(2024)\citenamefont {Manetsch}, \citenamefont {Nomura}, \citenamefont {Bataille}, \citenamefont {Leung}, \citenamefont {Lv},\ and\ \citenamefont {Endres}}]{manetsch2024tweezer}%
  \BibitemOpen
  \bibfield  {author} {\bibinfo {author} {\bibfnamefont {H.~J.}\ \bibnamefont {Manetsch}}, \bibinfo {author} {\bibfnamefont {G.}~\bibnamefont {Nomura}}, \bibinfo {author} {\bibfnamefont {E.}~\bibnamefont {Bataille}}, \bibinfo {author} {\bibfnamefont {K.~H.}\ \bibnamefont {Leung}}, \bibinfo {author} {\bibfnamefont {X.}~\bibnamefont {Lv}}, \ and\ \bibinfo {author} {\bibfnamefont {M.}~\bibnamefont {Endres}},\ }\href@noop {} {\bibfield  {journal} {\bibinfo  {journal} {arXiv preprint arXiv:2403.12021}\ } (\bibinfo {year} {2024})}\BibitemShut {NoStop}%
\bibitem [{\citenamefont {Chiu}\ \emph {et~al.}(2025)\citenamefont {Chiu}, \citenamefont {Trapp}, \citenamefont {Guo},\ and\ \citenamefont {Mohamed}}]{Chiu2025}%
  \BibitemOpen
  \bibfield  {author} {\bibinfo {author} {\bibfnamefont {N.-C.}\ \bibnamefont {Chiu}}, \bibinfo {author} {\bibfnamefont {E.~C.}\ \bibnamefont {Trapp}}, \bibinfo {author} {\bibfnamefont {J.}~\bibnamefont {Guo}}, \ and\ \bibinfo {author} {\bibfnamefont {H.}~\bibnamefont {Mohamed}},\ }\href {https://arxiv.org/abs/2506.20660} {\bibfield  {journal} {\bibinfo  {journal} {arXiv preprint arXiv:2506.20660}\ } (\bibinfo {year} {2025})},\ \Eprint {http://arxiv.org/abs/2506.20660} {arXiv:2506.20660 [quant-ph]} \BibitemShut {NoStop}%
\bibitem [{\citenamefont {Endres}\ \emph {et~al.}(2016)\citenamefont {Endres}, \citenamefont {Bernien}, \citenamefont {Keesling}, \citenamefont {Levine}, \citenamefont {Anschuetz}, \citenamefont {Krajenbrink}, \citenamefont {Senko}, \citenamefont {Vuletic}, \citenamefont {Greiner},\ and\ \citenamefont {Lukin}}]{endres2016atom}%
  \BibitemOpen
  \bibfield  {author} {\bibinfo {author} {\bibfnamefont {M.}~\bibnamefont {Endres}}, \bibinfo {author} {\bibfnamefont {H.}~\bibnamefont {Bernien}}, \bibinfo {author} {\bibfnamefont {A.}~\bibnamefont {Keesling}}, \bibinfo {author} {\bibfnamefont {H.}~\bibnamefont {Levine}}, \bibinfo {author} {\bibfnamefont {E.~R.}\ \bibnamefont {Anschuetz}}, \bibinfo {author} {\bibfnamefont {A.}~\bibnamefont {Krajenbrink}}, \bibinfo {author} {\bibfnamefont {C.}~\bibnamefont {Senko}}, \bibinfo {author} {\bibfnamefont {V.}~\bibnamefont {Vuletic}}, \bibinfo {author} {\bibfnamefont {M.}~\bibnamefont {Greiner}}, \ and\ \bibinfo {author} {\bibfnamefont {M.~D.}\ \bibnamefont {Lukin}},\ }\href@noop {} {\bibfield  {journal} {\bibinfo  {journal} {Science}\ }\textbf {\bibinfo {volume} {354}},\ \bibinfo {pages} {1024} (\bibinfo {year} {2016})}\BibitemShut {NoStop}%
\bibitem [{\citenamefont {Ma}\ \emph {et~al.}(2023)\citenamefont {Ma}, \citenamefont {Liu}, \citenamefont {Peng}, \citenamefont {Zhang}, \citenamefont {Jandura}, \citenamefont {Claes}, \citenamefont {Burgers}, \citenamefont {Pupillo}, \citenamefont {Puri},\ and\ \citenamefont {Thompson}}]{ma2023high}%
  \BibitemOpen
  \bibfield  {author} {\bibinfo {author} {\bibfnamefont {S.}~\bibnamefont {Ma}}, \bibinfo {author} {\bibfnamefont {G.}~\bibnamefont {Liu}}, \bibinfo {author} {\bibfnamefont {P.}~\bibnamefont {Peng}}, \bibinfo {author} {\bibfnamefont {B.}~\bibnamefont {Zhang}}, \bibinfo {author} {\bibfnamefont {S.}~\bibnamefont {Jandura}}, \bibinfo {author} {\bibfnamefont {J.}~\bibnamefont {Claes}}, \bibinfo {author} {\bibfnamefont {A.~P.}\ \bibnamefont {Burgers}}, \bibinfo {author} {\bibfnamefont {G.}~\bibnamefont {Pupillo}}, \bibinfo {author} {\bibfnamefont {S.}~\bibnamefont {Puri}}, \ and\ \bibinfo {author} {\bibfnamefont {J.~D.}\ \bibnamefont {Thompson}},\ }\href@noop {} {\bibfield  {journal} {\bibinfo  {journal} {Nature}\ }\textbf {\bibinfo {volume} {622}},\ \bibinfo {pages} {279} (\bibinfo {year} {2023})}\BibitemShut {NoStop}%
\bibitem [{\citenamefont {Evered}\ \emph {et~al.}(2023)\citenamefont {Evered}, \citenamefont {Bluvstein}, \citenamefont {Kalinowski}, \citenamefont {Ebadi}, \citenamefont {Manovitz}, \citenamefont {Zhou}, \citenamefont {Li}, \citenamefont {Geim}, \citenamefont {Wang}, \citenamefont {Maskara} \emph {et~al.}}]{evered2023high}%
  \BibitemOpen
  \bibfield  {author} {\bibinfo {author} {\bibfnamefont {S.~J.}\ \bibnamefont {Evered}}, \bibinfo {author} {\bibfnamefont {D.}~\bibnamefont {Bluvstein}}, \bibinfo {author} {\bibfnamefont {M.}~\bibnamefont {Kalinowski}}, \bibinfo {author} {\bibfnamefont {S.}~\bibnamefont {Ebadi}}, \bibinfo {author} {\bibfnamefont {T.}~\bibnamefont {Manovitz}}, \bibinfo {author} {\bibfnamefont {H.}~\bibnamefont {Zhou}}, \bibinfo {author} {\bibfnamefont {S.~H.}\ \bibnamefont {Li}}, \bibinfo {author} {\bibfnamefont {A.~A.}\ \bibnamefont {Geim}}, \bibinfo {author} {\bibfnamefont {T.~T.}\ \bibnamefont {Wang}}, \bibinfo {author} {\bibfnamefont {N.}~\bibnamefont {Maskara}},  \emph {et~al.},\ }\href@noop {} {\bibfield  {journal} {\bibinfo  {journal} {Nature}\ }\textbf {\bibinfo {volume} {622}},\ \bibinfo {pages} {268} (\bibinfo {year} {2023})}\BibitemShut {NoStop}%
\bibitem [{\citenamefont {Debnath}\ \emph {et~al.}(2016)\citenamefont {Debnath}, \citenamefont {Linke}, \citenamefont {Figgatt}, \citenamefont {Landsman}, \citenamefont {Wright},\ and\ \citenamefont {Monroe}}]{Debnath2016-sy}%
  \BibitemOpen
  \bibfield  {author} {\bibinfo {author} {\bibfnamefont {S.}~\bibnamefont {Debnath}}, \bibinfo {author} {\bibfnamefont {N.~M.}\ \bibnamefont {Linke}}, \bibinfo {author} {\bibfnamefont {C.}~\bibnamefont {Figgatt}}, \bibinfo {author} {\bibfnamefont {K.~A.}\ \bibnamefont {Landsman}}, \bibinfo {author} {\bibfnamefont {K.}~\bibnamefont {Wright}}, \ and\ \bibinfo {author} {\bibfnamefont {C.}~\bibnamefont {Monroe}},\ }\href@noop {} {\bibfield  {journal} {\bibinfo  {journal} {Nature}\ }\textbf {\bibinfo {volume} {536}},\ \bibinfo {pages} {63} (\bibinfo {year} {2016})}\BibitemShut {NoStop}%
\bibitem [{\citenamefont {Zhang}\ \emph {et~al.}(2024)\citenamefont {Zhang}, \citenamefont {Peng}, \citenamefont {Paul},\ and\ \citenamefont {Thompson}}]{zhang2024scaled}%
  \BibitemOpen
  \bibfield  {author} {\bibinfo {author} {\bibfnamefont {B.}~\bibnamefont {Zhang}}, \bibinfo {author} {\bibfnamefont {P.}~\bibnamefont {Peng}}, \bibinfo {author} {\bibfnamefont {A.}~\bibnamefont {Paul}}, \ and\ \bibinfo {author} {\bibfnamefont {J.~D.}\ \bibnamefont {Thompson}},\ }\href@noop {} {\bibfield  {journal} {\bibinfo  {journal} {Optica}\ }\textbf {\bibinfo {volume} {11}},\ \bibinfo {pages} {227} (\bibinfo {year} {2024})}\BibitemShut {NoStop}%
\bibitem [{\citenamefont {Mehta}\ \emph {et~al.}(2016)\citenamefont {Mehta}, \citenamefont {Bruzewicz}, \citenamefont {McConnell}, \citenamefont {Ram}, \citenamefont {Sage},\ and\ \citenamefont {Chiaverini}}]{Mehta2016-ak}%
  \BibitemOpen
  \bibfield  {author} {\bibinfo {author} {\bibfnamefont {K.~K.}\ \bibnamefont {Mehta}}, \bibinfo {author} {\bibfnamefont {C.~D.}\ \bibnamefont {Bruzewicz}}, \bibinfo {author} {\bibfnamefont {R.}~\bibnamefont {McConnell}}, \bibinfo {author} {\bibfnamefont {R.~J.}\ \bibnamefont {Ram}}, \bibinfo {author} {\bibfnamefont {J.~M.}\ \bibnamefont {Sage}}, \ and\ \bibinfo {author} {\bibfnamefont {J.}~\bibnamefont {Chiaverini}},\ }\href@noop {} {\bibfield  {journal} {\bibinfo  {journal} {Nature Nanotechnology}\ }\textbf {\bibinfo {volume} {11}},\ \bibinfo {pages} {1066} (\bibinfo {year} {2016})}\BibitemShut {NoStop}%
\bibitem [{\citenamefont {Newman}\ \emph {et~al.}(2019)\citenamefont {Newman}, \citenamefont {Maurice}, \citenamefont {Drake}, \citenamefont {Stone}, \citenamefont {Briles}, \citenamefont {Spencer}, \citenamefont {Fredrick}, \citenamefont {Li}, \citenamefont {Westly}, \citenamefont {Ilic} \emph {et~al.}}]{newman2019architecture}%
  \BibitemOpen
  \bibfield  {author} {\bibinfo {author} {\bibfnamefont {Z.~L.}\ \bibnamefont {Newman}}, \bibinfo {author} {\bibfnamefont {V.}~\bibnamefont {Maurice}}, \bibinfo {author} {\bibfnamefont {T.}~\bibnamefont {Drake}}, \bibinfo {author} {\bibfnamefont {J.~R.}\ \bibnamefont {Stone}}, \bibinfo {author} {\bibfnamefont {T.~C.}\ \bibnamefont {Briles}}, \bibinfo {author} {\bibfnamefont {D.~T.}\ \bibnamefont {Spencer}}, \bibinfo {author} {\bibfnamefont {C.}~\bibnamefont {Fredrick}}, \bibinfo {author} {\bibfnamefont {Q.}~\bibnamefont {Li}}, \bibinfo {author} {\bibfnamefont {D.}~\bibnamefont {Westly}}, \bibinfo {author} {\bibfnamefont {B.~R.}\ \bibnamefont {Ilic}},  \emph {et~al.},\ }\href@noop {} {\bibfield  {journal} {\bibinfo  {journal} {Optica}\ }\textbf {\bibinfo {volume} {6}},\ \bibinfo {pages} {680} (\bibinfo {year} {2019})}\BibitemShut {NoStop}%
\bibitem [{\citenamefont {Niffenegger}\ \emph {et~al.}(2020)\citenamefont {Niffenegger}, \citenamefont {Stuart}, \citenamefont {Sorace-Agaskar}, \citenamefont {Kharas}, \citenamefont {Bramhavar}, \citenamefont {Bruzewicz}, \citenamefont {Loh}, \citenamefont {Maxson}, \citenamefont {McConnell}, \citenamefont {Reens}, \citenamefont {West}, \citenamefont {Sage},\ and\ \citenamefont {Chiaverini}}]{Niffenegger2020-up}%
  \BibitemOpen
  \bibfield  {author} {\bibinfo {author} {\bibfnamefont {R.~J.}\ \bibnamefont {Niffenegger}}, \bibinfo {author} {\bibfnamefont {J.}~\bibnamefont {Stuart}}, \bibinfo {author} {\bibfnamefont {C.}~\bibnamefont {Sorace-Agaskar}}, \bibinfo {author} {\bibfnamefont {D.}~\bibnamefont {Kharas}}, \bibinfo {author} {\bibfnamefont {S.}~\bibnamefont {Bramhavar}}, \bibinfo {author} {\bibfnamefont {C.~D.}\ \bibnamefont {Bruzewicz}}, \bibinfo {author} {\bibfnamefont {W.}~\bibnamefont {Loh}}, \bibinfo {author} {\bibfnamefont {R.~T.}\ \bibnamefont {Maxson}}, \bibinfo {author} {\bibfnamefont {R.}~\bibnamefont {McConnell}}, \bibinfo {author} {\bibfnamefont {D.}~\bibnamefont {Reens}}, \bibinfo {author} {\bibfnamefont {G.~N.}\ \bibnamefont {West}}, \bibinfo {author} {\bibfnamefont {J.~M.}\ \bibnamefont {Sage}}, \ and\ \bibinfo {author} {\bibfnamefont {J.}~\bibnamefont {Chiaverini}},\ }\href@noop {} {\bibfield  {journal} {\bibinfo  {journal} {Nature}\ }\textbf {\bibinfo {volume} {586}},\ \bibinfo {pages} {538} (\bibinfo {year}
  {2020})}\BibitemShut {NoStop}%
\bibitem [{\citenamefont {Mehta}\ \emph {et~al.}(2020)\citenamefont {Mehta}, \citenamefont {Zhang}, \citenamefont {Malinowski}, \citenamefont {Nguyen}, \citenamefont {Stadler},\ and\ \citenamefont {Home}}]{Mehta2020-in}%
  \BibitemOpen
  \bibfield  {author} {\bibinfo {author} {\bibfnamefont {K.~K.}\ \bibnamefont {Mehta}}, \bibinfo {author} {\bibfnamefont {C.}~\bibnamefont {Zhang}}, \bibinfo {author} {\bibfnamefont {M.}~\bibnamefont {Malinowski}}, \bibinfo {author} {\bibfnamefont {T.-L.}\ \bibnamefont {Nguyen}}, \bibinfo {author} {\bibfnamefont {M.}~\bibnamefont {Stadler}}, \ and\ \bibinfo {author} {\bibfnamefont {J.~P.}\ \bibnamefont {Home}},\ }\href@noop {} {\bibfield  {journal} {\bibinfo  {journal} {Nature}\ }\textbf {\bibinfo {volume} {586}},\ \bibinfo {pages} {533} (\bibinfo {year} {2020})}\BibitemShut {NoStop}%
\bibitem [{\citenamefont {Wan}\ \emph {et~al.}(2020)\citenamefont {Wan}, \citenamefont {Lu}, \citenamefont {Chen}, \citenamefont {Walsh}, \citenamefont {Trusheim}, \citenamefont {De~Santis}, \citenamefont {Bersin}, \citenamefont {Harris}, \citenamefont {Mouradian}, \citenamefont {Christen}, \citenamefont {Bielejec},\ and\ \citenamefont {Englund}}]{Wan2020-xs}%
  \BibitemOpen
  \bibfield  {author} {\bibinfo {author} {\bibfnamefont {N.~H.}\ \bibnamefont {Wan}}, \bibinfo {author} {\bibfnamefont {T.-J.}\ \bibnamefont {Lu}}, \bibinfo {author} {\bibfnamefont {K.~C.}\ \bibnamefont {Chen}}, \bibinfo {author} {\bibfnamefont {M.~P.}\ \bibnamefont {Walsh}}, \bibinfo {author} {\bibfnamefont {M.~E.}\ \bibnamefont {Trusheim}}, \bibinfo {author} {\bibfnamefont {L.}~\bibnamefont {De~Santis}}, \bibinfo {author} {\bibfnamefont {E.~A.}\ \bibnamefont {Bersin}}, \bibinfo {author} {\bibfnamefont {I.~B.}\ \bibnamefont {Harris}}, \bibinfo {author} {\bibfnamefont {S.~L.}\ \bibnamefont {Mouradian}}, \bibinfo {author} {\bibfnamefont {I.~R.}\ \bibnamefont {Christen}}, \bibinfo {author} {\bibfnamefont {E.~S.}\ \bibnamefont {Bielejec}}, \ and\ \bibinfo {author} {\bibfnamefont {D.}~\bibnamefont {Englund}},\ }\href@noop {} {\bibfield  {journal} {\bibinfo  {journal} {Nature}\ }\textbf {\bibinfo {volume} {583}},\ \bibinfo {pages} {226} (\bibinfo {year} {2020})}\BibitemShut {NoStop}%
\bibitem [{\citenamefont {Hogle}\ \emph {et~al.}(2023)\citenamefont {Hogle}, \citenamefont {Dominguez}, \citenamefont {Dong}, \citenamefont {Leenheer}, \citenamefont {McGuinness}, \citenamefont {Ruzic}, \citenamefont {Eichenfield},\ and\ \citenamefont {Stick}}]{hogle2023high}%
  \BibitemOpen
  \bibfield  {author} {\bibinfo {author} {\bibfnamefont {C.~W.}\ \bibnamefont {Hogle}}, \bibinfo {author} {\bibfnamefont {D.}~\bibnamefont {Dominguez}}, \bibinfo {author} {\bibfnamefont {M.}~\bibnamefont {Dong}}, \bibinfo {author} {\bibfnamefont {A.}~\bibnamefont {Leenheer}}, \bibinfo {author} {\bibfnamefont {H.~J.}\ \bibnamefont {McGuinness}}, \bibinfo {author} {\bibfnamefont {B.~P.}\ \bibnamefont {Ruzic}}, \bibinfo {author} {\bibfnamefont {M.}~\bibnamefont {Eichenfield}}, \ and\ \bibinfo {author} {\bibfnamefont {D.}~\bibnamefont {Stick}},\ }\href@noop {} {\bibfield  {journal} {\bibinfo  {journal} {npj Quantum Information}\ }\textbf {\bibinfo {volume} {9}},\ \bibinfo {pages} {74} (\bibinfo {year} {2023})}\BibitemShut {NoStop}%
\bibitem [{\citenamefont {Isichenko}\ \emph {et~al.}(2023)\citenamefont {Isichenko}, \citenamefont {Chauhan}, \citenamefont {Bose}, \citenamefont {Wang}, \citenamefont {Kunz},\ and\ \citenamefont {Blumenthal}}]{Isichenko2023-gr}%
  \BibitemOpen
  \bibfield  {author} {\bibinfo {author} {\bibfnamefont {A.}~\bibnamefont {Isichenko}}, \bibinfo {author} {\bibfnamefont {N.}~\bibnamefont {Chauhan}}, \bibinfo {author} {\bibfnamefont {D.}~\bibnamefont {Bose}}, \bibinfo {author} {\bibfnamefont {J.}~\bibnamefont {Wang}}, \bibinfo {author} {\bibfnamefont {P.~D.}\ \bibnamefont {Kunz}}, \ and\ \bibinfo {author} {\bibfnamefont {D.~J.}\ \bibnamefont {Blumenthal}},\ }\href@noop {} {\bibfield  {journal} {\bibinfo  {journal} {Nature Communications}\ }\textbf {\bibinfo {volume} {14}},\ \bibinfo {pages} {3080} (\bibinfo {year} {2023})}\BibitemShut {NoStop}%
\bibitem [{\citenamefont {Knollmann}\ \emph {et~al.}(2024)\citenamefont {Knollmann}, \citenamefont {Clements}, \citenamefont {Callahan}, \citenamefont {Gehl}, \citenamefont {Hunker}, \citenamefont {Mahony}, \citenamefont {McConnell}, \citenamefont {Swint}, \citenamefont {Sorace-Agaskar}, \citenamefont {Chuang}, \citenamefont {Chiaverini},\ and\ \citenamefont {Stick}}]{Knollmann:24}%
  \BibitemOpen
  \bibfield  {author} {\bibinfo {author} {\bibfnamefont {F.~W.}\ \bibnamefont {Knollmann}}, \bibinfo {author} {\bibfnamefont {E.}~\bibnamefont {Clements}}, \bibinfo {author} {\bibfnamefont {P.~T.}\ \bibnamefont {Callahan}}, \bibinfo {author} {\bibfnamefont {M.}~\bibnamefont {Gehl}}, \bibinfo {author} {\bibfnamefont {J.~D.}\ \bibnamefont {Hunker}}, \bibinfo {author} {\bibfnamefont {T.}~\bibnamefont {Mahony}}, \bibinfo {author} {\bibfnamefont {R.}~\bibnamefont {McConnell}}, \bibinfo {author} {\bibfnamefont {R.}~\bibnamefont {Swint}}, \bibinfo {author} {\bibfnamefont {C.}~\bibnamefont {Sorace-Agaskar}}, \bibinfo {author} {\bibfnamefont {I.~L.}\ \bibnamefont {Chuang}}, \bibinfo {author} {\bibfnamefont {J.}~\bibnamefont {Chiaverini}}, \ and\ \bibinfo {author} {\bibfnamefont {D.}~\bibnamefont {Stick}},\ }\href {\doibase 10.1364/OPTICAQ.522128} {\bibfield  {journal} {\bibinfo  {journal} {Optica Quantum}\ }\textbf {\bibinfo {volume} {2}},\ \bibinfo {pages} {230} (\bibinfo {year} {2024})}\BibitemShut {NoStop}%
\bibitem [{\citenamefont {Menon}\ \emph {et~al.}(2024)\citenamefont {Menon}, \citenamefont {Glachman}, \citenamefont {Pompili}, \citenamefont {Dibos},\ and\ \citenamefont {Bernien}}]{Menon2024-lo}%
  \BibitemOpen
  \bibfield  {author} {\bibinfo {author} {\bibfnamefont {S.~G.}\ \bibnamefont {Menon}}, \bibinfo {author} {\bibfnamefont {N.}~\bibnamefont {Glachman}}, \bibinfo {author} {\bibfnamefont {M.}~\bibnamefont {Pompili}}, \bibinfo {author} {\bibfnamefont {A.}~\bibnamefont {Dibos}}, \ and\ \bibinfo {author} {\bibfnamefont {H.}~\bibnamefont {Bernien}},\ }\href@noop {} {\bibfield  {journal} {\bibinfo  {journal} {Nature Communications}\ }\textbf {\bibinfo {volume} {15}},\ \bibinfo {pages} {6156} (\bibinfo {year} {2024})}\BibitemShut {NoStop}%
\bibitem [{\citenamefont {Craft}\ \emph {et~al.}(2024)\citenamefont {Craft}, \citenamefont {Barton}, \citenamefont {Klug}, \citenamefont {Scalzi}, \citenamefont {Wildemann}, \citenamefont {Asagodu}, \citenamefont {Broz}, \citenamefont {Porto}, \citenamefont {Macalik}, \citenamefont {Rizzo}, \citenamefont {Percevault}, \citenamefont {Tison}, \citenamefont {Smith}, \citenamefont {Fanto}, \citenamefont {Schneeloch}, \citenamefont {Sheridan}, \citenamefont {Heberle}, \citenamefont {Brownell}, \citenamefont {Sundaram}, \citenamefont {Deenadayalan}, \citenamefont {van Niekerk}, \citenamefont {Manfreda-Schulz}, \citenamefont {Howland}, \citenamefont {Preble}, \citenamefont {Coleman}, \citenamefont {Leake}, \citenamefont {Antohe}, \citenamefont {Vo}, \citenamefont {Fahrenkopf}, \citenamefont {Stievater}, \citenamefont {Brickman-Soderberg}, \citenamefont {Smith},\ and\ \citenamefont {Hucul}}]{craft2024low}%
  \BibitemOpen
  \bibfield  {author} {\bibinfo {author} {\bibfnamefont {C.~L.}\ \bibnamefont {Craft}}, \bibinfo {author} {\bibfnamefont {N.~J.}\ \bibnamefont {Barton}}, \bibinfo {author} {\bibfnamefont {A.~C.}\ \bibnamefont {Klug}}, \bibinfo {author} {\bibfnamefont {K.}~\bibnamefont {Scalzi}}, \bibinfo {author} {\bibfnamefont {I.}~\bibnamefont {Wildemann}}, \bibinfo {author} {\bibfnamefont {P.}~\bibnamefont {Asagodu}}, \bibinfo {author} {\bibfnamefont {J.~D.}\ \bibnamefont {Broz}}, \bibinfo {author} {\bibfnamefont {N.~L.}\ \bibnamefont {Porto}}, \bibinfo {author} {\bibfnamefont {M.}~\bibnamefont {Macalik}}, \bibinfo {author} {\bibfnamefont {A.}~\bibnamefont {Rizzo}}, \bibinfo {author} {\bibfnamefont {G.}~\bibnamefont {Percevault}}, \bibinfo {author} {\bibfnamefont {C.~C.}\ \bibnamefont {Tison}}, \bibinfo {author} {\bibfnamefont {A.~M.}\ \bibnamefont {Smith}}, \bibinfo {author} {\bibfnamefont {M.~L.}\ \bibnamefont {Fanto}}, \bibinfo {author} {\bibfnamefont {J.}~\bibnamefont {Schneeloch}}, \bibinfo {author} {\bibfnamefont
  {E.}~\bibnamefont {Sheridan}}, \bibinfo {author} {\bibfnamefont {D.}~\bibnamefont {Heberle}}, \bibinfo {author} {\bibfnamefont {A.}~\bibnamefont {Brownell}}, \bibinfo {author} {\bibfnamefont {V.~S.~S.}\ \bibnamefont {Sundaram}}, \bibinfo {author} {\bibfnamefont {V.}~\bibnamefont {Deenadayalan}}, \bibinfo {author} {\bibfnamefont {M.}~\bibnamefont {van Niekerk}}, \bibinfo {author} {\bibfnamefont {E.}~\bibnamefont {Manfreda-Schulz}}, \bibinfo {author} {\bibfnamefont {G.~A.}\ \bibnamefont {Howland}}, \bibinfo {author} {\bibfnamefont {S.~F.}\ \bibnamefont {Preble}}, \bibinfo {author} {\bibfnamefont {D.}~\bibnamefont {Coleman}}, \bibinfo {author} {\bibfnamefont {G.}~\bibnamefont {Leake}}, \bibinfo {author} {\bibfnamefont {A.}~\bibnamefont {Antohe}}, \bibinfo {author} {\bibfnamefont {T.}~\bibnamefont {Vo}}, \bibinfo {author} {\bibfnamefont {N.~M.}\ \bibnamefont {Fahrenkopf}}, \bibinfo {author} {\bibfnamefont {T.~H.}\ \bibnamefont {Stievater}}, \bibinfo {author} {\bibfnamefont {K.-A.}\ \bibnamefont
  {Brickman-Soderberg}}, \bibinfo {author} {\bibfnamefont {Z.~S.}\ \bibnamefont {Smith}}, \ and\ \bibinfo {author} {\bibfnamefont {D.}~\bibnamefont {Hucul}},\ }\href {https://arxiv.org/abs/2406.17607} {\enquote {\bibinfo {title} {Low-crosstalk, silicon-fabricated optical waveguides for laser delivery to matter qubits},}\ } (\bibinfo {year} {2024}),\ \Eprint {http://arxiv.org/abs/2406.17607} {arXiv:2406.17607 [quant-ph]} \BibitemShut {NoStop}%
\bibitem [{\citenamefont {Loh}\ \emph {et~al.}(2025)\citenamefont {Loh}, \citenamefont {Reens}, \citenamefont {Kharas}, \citenamefont {Sumant}, \citenamefont {Belanger}, \citenamefont {Maxson}, \citenamefont {Medeiros}, \citenamefont {Setzer}, \citenamefont {Gray}, \citenamefont {DeBry}, \citenamefont {Bruzewicz}, \citenamefont {Plant}, \citenamefont {Liddell}, \citenamefont {West}, \citenamefont {Doshi}, \citenamefont {Roychowdhury}, \citenamefont {Kim}, \citenamefont {Braje}, \citenamefont {Juodawlkis}, \citenamefont {Chiaverini},\ and\ \citenamefont {McConnell}}]{Loh2025-ci}%
  \BibitemOpen
  \bibfield  {author} {\bibinfo {author} {\bibfnamefont {W.}~\bibnamefont {Loh}}, \bibinfo {author} {\bibfnamefont {D.}~\bibnamefont {Reens}}, \bibinfo {author} {\bibfnamefont {D.}~\bibnamefont {Kharas}}, \bibinfo {author} {\bibfnamefont {A.}~\bibnamefont {Sumant}}, \bibinfo {author} {\bibfnamefont {C.}~\bibnamefont {Belanger}}, \bibinfo {author} {\bibfnamefont {R.~T.}\ \bibnamefont {Maxson}}, \bibinfo {author} {\bibfnamefont {A.}~\bibnamefont {Medeiros}}, \bibinfo {author} {\bibfnamefont {W.}~\bibnamefont {Setzer}}, \bibinfo {author} {\bibfnamefont {D.}~\bibnamefont {Gray}}, \bibinfo {author} {\bibfnamefont {K.}~\bibnamefont {DeBry}}, \bibinfo {author} {\bibfnamefont {C.~D.}\ \bibnamefont {Bruzewicz}}, \bibinfo {author} {\bibfnamefont {J.}~\bibnamefont {Plant}}, \bibinfo {author} {\bibfnamefont {J.}~\bibnamefont {Liddell}}, \bibinfo {author} {\bibfnamefont {G.~N.}\ \bibnamefont {West}}, \bibinfo {author} {\bibfnamefont {S.}~\bibnamefont {Doshi}}, \bibinfo {author} {\bibfnamefont {M.}~\bibnamefont
  {Roychowdhury}}, \bibinfo {author} {\bibfnamefont {M.~E.}\ \bibnamefont {Kim}}, \bibinfo {author} {\bibfnamefont {D.}~\bibnamefont {Braje}}, \bibinfo {author} {\bibfnamefont {P.~W.}\ \bibnamefont {Juodawlkis}}, \bibinfo {author} {\bibfnamefont {J.}~\bibnamefont {Chiaverini}}, \ and\ \bibinfo {author} {\bibfnamefont {R.}~\bibnamefont {McConnell}},\ }\href@noop {} {\bibfield  {journal} {\bibinfo  {journal} {Nature Photonics}\ }\textbf {\bibinfo {volume} {19}},\ \bibinfo {pages} {277} (\bibinfo {year} {2025})}\BibitemShut {NoStop}%
\bibitem [{\citenamefont {Menssen}\ \emph {et~al.}(2023)\citenamefont {Menssen}, \citenamefont {Hermans}, \citenamefont {Christen}, \citenamefont {Propson}, \citenamefont {Li}, \citenamefont {Leenheer}, \citenamefont {Zimmermann}, \citenamefont {Dong}, \citenamefont {Larocque}, \citenamefont {Raniwala}, \citenamefont {Gilbert}, \citenamefont {Eichenfield},\ and\ \citenamefont {Englund}}]{Menssen:2_optica}%
  \BibitemOpen
  \bibfield  {author} {\bibinfo {author} {\bibfnamefont {A.~J.}\ \bibnamefont {Menssen}}, \bibinfo {author} {\bibfnamefont {A.}~\bibnamefont {Hermans}}, \bibinfo {author} {\bibfnamefont {I.}~\bibnamefont {Christen}}, \bibinfo {author} {\bibfnamefont {T.}~\bibnamefont {Propson}}, \bibinfo {author} {\bibfnamefont {C.}~\bibnamefont {Li}}, \bibinfo {author} {\bibfnamefont {A.~J.}\ \bibnamefont {Leenheer}}, \bibinfo {author} {\bibfnamefont {M.}~\bibnamefont {Zimmermann}}, \bibinfo {author} {\bibfnamefont {M.}~\bibnamefont {Dong}}, \bibinfo {author} {\bibfnamefont {H.}~\bibnamefont {Larocque}}, \bibinfo {author} {\bibfnamefont {H.}~\bibnamefont {Raniwala}}, \bibinfo {author} {\bibfnamefont {G.}~\bibnamefont {Gilbert}}, \bibinfo {author} {\bibfnamefont {M.}~\bibnamefont {Eichenfield}}, \ and\ \bibinfo {author} {\bibfnamefont {D.~R.}\ \bibnamefont {Englund}},\ }\href {\doibase 10.1364/OPTICA.489504} {\bibfield  {journal} {\bibinfo  {journal} {Optica}\ }\textbf {\bibinfo {volume} {10}},\ \bibinfo {pages} {1366}
  (\bibinfo {year} {2023})}\BibitemShut {NoStop}%
\bibitem [{\citenamefont {Palm}\ \emph {et~al.}(2023)\citenamefont {Palm}, \citenamefont {Dong}, \citenamefont {Golter}, \citenamefont {Clark}, \citenamefont {Zimmermann}, \citenamefont {Chen}, \citenamefont {Li}, \citenamefont {Menssen}, \citenamefont {Leenheer}, \citenamefont {Dominguez}, \citenamefont {Gilbert}, \citenamefont {Eichenfield},\ and\ \citenamefont {Englund}}]{Palm2023-ju}%
  \BibitemOpen
  \bibfield  {author} {\bibinfo {author} {\bibfnamefont {K.~J.}\ \bibnamefont {Palm}}, \bibinfo {author} {\bibfnamefont {M.}~\bibnamefont {Dong}}, \bibinfo {author} {\bibfnamefont {D.~A.}\ \bibnamefont {Golter}}, \bibinfo {author} {\bibfnamefont {G.}~\bibnamefont {Clark}}, \bibinfo {author} {\bibfnamefont {M.}~\bibnamefont {Zimmermann}}, \bibinfo {author} {\bibfnamefont {K.~C.}\ \bibnamefont {Chen}}, \bibinfo {author} {\bibfnamefont {L.}~\bibnamefont {Li}}, \bibinfo {author} {\bibfnamefont {A.}~\bibnamefont {Menssen}}, \bibinfo {author} {\bibfnamefont {A.~J.}\ \bibnamefont {Leenheer}}, \bibinfo {author} {\bibfnamefont {D.}~\bibnamefont {Dominguez}}, \bibinfo {author} {\bibfnamefont {G.}~\bibnamefont {Gilbert}}, \bibinfo {author} {\bibfnamefont {M.}~\bibnamefont {Eichenfield}}, \ and\ \bibinfo {author} {\bibfnamefont {D.}~\bibnamefont {Englund}},\ }\href@noop {} {\bibfield  {journal} {\bibinfo  {journal} {Optica}\ }\textbf {\bibinfo {volume} {10}},\ \bibinfo {pages} {634} (\bibinfo {year} {2023})}\BibitemShut
  {NoStop}%
\bibitem [{\citenamefont {Christen}\ \emph {et~al.}(2025)\citenamefont {Christen}, \citenamefont {Propson}, \citenamefont {Sutula}, \citenamefont {Sattari}, \citenamefont {Choong}, \citenamefont {Panuski}, \citenamefont {Melville}, \citenamefont {Mallek}, \citenamefont {Brabec}, \citenamefont {Hamilton}, \citenamefont {Dixon}, \citenamefont {Menssen}, \citenamefont {Braje}, \citenamefont {Ghadimi},\ and\ \citenamefont {Englund}}]{Christen2025-cf}%
  \BibitemOpen
  \bibfield  {author} {\bibinfo {author} {\bibfnamefont {I.}~\bibnamefont {Christen}}, \bibinfo {author} {\bibfnamefont {T.}~\bibnamefont {Propson}}, \bibinfo {author} {\bibfnamefont {M.}~\bibnamefont {Sutula}}, \bibinfo {author} {\bibfnamefont {H.}~\bibnamefont {Sattari}}, \bibinfo {author} {\bibfnamefont {G.}~\bibnamefont {Choong}}, \bibinfo {author} {\bibfnamefont {C.}~\bibnamefont {Panuski}}, \bibinfo {author} {\bibfnamefont {A.}~\bibnamefont {Melville}}, \bibinfo {author} {\bibfnamefont {J.}~\bibnamefont {Mallek}}, \bibinfo {author} {\bibfnamefont {C.}~\bibnamefont {Brabec}}, \bibinfo {author} {\bibfnamefont {S.}~\bibnamefont {Hamilton}}, \bibinfo {author} {\bibfnamefont {P.~B.}\ \bibnamefont {Dixon}}, \bibinfo {author} {\bibfnamefont {A.~J.}\ \bibnamefont {Menssen}}, \bibinfo {author} {\bibfnamefont {D.}~\bibnamefont {Braje}}, \bibinfo {author} {\bibfnamefont {A.~H.}\ \bibnamefont {Ghadimi}}, \ and\ \bibinfo {author} {\bibfnamefont {D.}~\bibnamefont {Englund}},\ }\href@noop {} {\bibfield  {journal}
  {\bibinfo  {journal} {Nature Communications}\ }\textbf {\bibinfo {volume} {16}},\ \bibinfo {pages} {82} (\bibinfo {year} {2025})}\BibitemShut {NoStop}%
\bibitem [{\citenamefont {Tran}\ \emph {et~al.}(2022)\citenamefont {Tran}, \citenamefont {Zhang}, \citenamefont {Morin}, \citenamefont {Chang}, \citenamefont {Barik}, \citenamefont {Yuan}, \citenamefont {Lee}, \citenamefont {Kim}, \citenamefont {Malik}, \citenamefont {Zhang} \emph {et~al.}}]{tran2022extending}%
  \BibitemOpen
  \bibfield  {author} {\bibinfo {author} {\bibfnamefont {M.~A.}\ \bibnamefont {Tran}}, \bibinfo {author} {\bibfnamefont {C.}~\bibnamefont {Zhang}}, \bibinfo {author} {\bibfnamefont {T.~J.}\ \bibnamefont {Morin}}, \bibinfo {author} {\bibfnamefont {L.}~\bibnamefont {Chang}}, \bibinfo {author} {\bibfnamefont {S.}~\bibnamefont {Barik}}, \bibinfo {author} {\bibfnamefont {Z.}~\bibnamefont {Yuan}}, \bibinfo {author} {\bibfnamefont {W.}~\bibnamefont {Lee}}, \bibinfo {author} {\bibfnamefont {G.}~\bibnamefont {Kim}}, \bibinfo {author} {\bibfnamefont {A.}~\bibnamefont {Malik}}, \bibinfo {author} {\bibfnamefont {Z.}~\bibnamefont {Zhang}},  \emph {et~al.},\ }\href@noop {} {\bibfield  {journal} {\bibinfo  {journal} {Nature}\ }\textbf {\bibinfo {volume} {610}},\ \bibinfo {pages} {54} (\bibinfo {year} {2022})}\BibitemShut {NoStop}%
\bibitem [{\citenamefont {Corato-Zanarella}\ \emph {et~al.}(2023)\citenamefont {Corato-Zanarella}, \citenamefont {Gil-Molina}, \citenamefont {Ji}, \citenamefont {Shin}, \citenamefont {Mohanty},\ and\ \citenamefont {Lipson}}]{Corato-Zanarella2023-bo}%
  \BibitemOpen
  \bibfield  {author} {\bibinfo {author} {\bibfnamefont {M.}~\bibnamefont {Corato-Zanarella}}, \bibinfo {author} {\bibfnamefont {A.}~\bibnamefont {Gil-Molina}}, \bibinfo {author} {\bibfnamefont {X.}~\bibnamefont {Ji}}, \bibinfo {author} {\bibfnamefont {M.~C.}\ \bibnamefont {Shin}}, \bibinfo {author} {\bibfnamefont {A.}~\bibnamefont {Mohanty}}, \ and\ \bibinfo {author} {\bibfnamefont {M.}~\bibnamefont {Lipson}},\ }\href@noop {} {\bibfield  {journal} {\bibinfo  {journal} {Nature Photonics}\ }\textbf {\bibinfo {volume} {17}},\ \bibinfo {pages} {157} (\bibinfo {year} {2023})}\BibitemShut {NoStop}%
\bibitem [{\citenamefont {Park}\ \emph {et~al.}(2024)\citenamefont {Park}, \citenamefont {Notaros}, \citenamefont {Mohanty}, \citenamefont {Kim}, \citenamefont {Notaros},\ and\ \citenamefont {Mouradian}}]{park2024technologies}%
  \BibitemOpen
  \bibfield  {author} {\bibinfo {author} {\bibfnamefont {S.}~\bibnamefont {Park}}, \bibinfo {author} {\bibfnamefont {M.}~\bibnamefont {Notaros}}, \bibinfo {author} {\bibfnamefont {A.}~\bibnamefont {Mohanty}}, \bibinfo {author} {\bibfnamefont {D.}~\bibnamefont {Kim}}, \bibinfo {author} {\bibfnamefont {J.}~\bibnamefont {Notaros}}, \ and\ \bibinfo {author} {\bibfnamefont {S.}~\bibnamefont {Mouradian}},\ }\href@noop {} {\bibfield  {journal} {\bibinfo  {journal} {Progress in Quantum Electronics}\ }\textbf {\bibinfo {volume} {97}},\ \bibinfo {pages} {100534} (\bibinfo {year} {2024})}\BibitemShut {NoStop}%
\bibitem [{\citenamefont {Kitaev}(2003)}]{kitaev2003fault}%
  \BibitemOpen
  \bibfield  {author} {\bibinfo {author} {\bibfnamefont {A.~Y.}\ \bibnamefont {Kitaev}},\ }\href@noop {} {\bibfield  {journal} {\bibinfo  {journal} {Annals of physics}\ }\textbf {\bibinfo {volume} {303}},\ \bibinfo {pages} {2} (\bibinfo {year} {2003})}\BibitemShut {NoStop}%
\bibitem [{\citenamefont {Fowler}\ \emph {et~al.}(2012)\citenamefont {Fowler}, \citenamefont {Mariantoni}, \citenamefont {Martinis},\ and\ \citenamefont {Cleland}}]{fowler2012surface}%
  \BibitemOpen
  \bibfield  {author} {\bibinfo {author} {\bibfnamefont {A.~G.}\ \bibnamefont {Fowler}}, \bibinfo {author} {\bibfnamefont {M.}~\bibnamefont {Mariantoni}}, \bibinfo {author} {\bibfnamefont {J.~M.}\ \bibnamefont {Martinis}}, \ and\ \bibinfo {author} {\bibfnamefont {A.~N.}\ \bibnamefont {Cleland}},\ }\href@noop {} {\bibfield  {journal} {\bibinfo  {journal} {Physical Review A—Atomic, Molecular, and Optical Physics}\ }\textbf {\bibinfo {volume} {86}},\ \bibinfo {pages} {032324} (\bibinfo {year} {2012})}\BibitemShut {NoStop}%
\bibitem [{\citenamefont {Stanfield}\ \emph {et~al.}(2019)\citenamefont {Stanfield}, \citenamefont {Leenheer}, \citenamefont {Michael}, \citenamefont {Sims},\ and\ \citenamefont {Eichenfield}}]{stanfield2019cmos}%
  \BibitemOpen
  \bibfield  {author} {\bibinfo {author} {\bibfnamefont {P.}~\bibnamefont {Stanfield}}, \bibinfo {author} {\bibfnamefont {A.}~\bibnamefont {Leenheer}}, \bibinfo {author} {\bibfnamefont {C.}~\bibnamefont {Michael}}, \bibinfo {author} {\bibfnamefont {R.}~\bibnamefont {Sims}}, \ and\ \bibinfo {author} {\bibfnamefont {M.}~\bibnamefont {Eichenfield}},\ }\href@noop {} {\bibfield  {journal} {\bibinfo  {journal} {Optics express}\ }\textbf {\bibinfo {volume} {27}},\ \bibinfo {pages} {28588} (\bibinfo {year} {2019})}\BibitemShut {NoStop}%
\bibitem [{\citenamefont {Dong}\ \emph {et~al.}(2022{\natexlab{a}})\citenamefont {Dong}, \citenamefont {Clark}, \citenamefont {Leenheer}, \citenamefont {Zimmermann}, \citenamefont {Dominguez}, \citenamefont {Menssen}, \citenamefont {Heim}, \citenamefont {Gilbert}, \citenamefont {Englund},\ and\ \citenamefont {Eichenfield}}]{dong2022high}%
  \BibitemOpen
  \bibfield  {author} {\bibinfo {author} {\bibfnamefont {M.}~\bibnamefont {Dong}}, \bibinfo {author} {\bibfnamefont {G.}~\bibnamefont {Clark}}, \bibinfo {author} {\bibfnamefont {A.~J.}\ \bibnamefont {Leenheer}}, \bibinfo {author} {\bibfnamefont {M.}~\bibnamefont {Zimmermann}}, \bibinfo {author} {\bibfnamefont {D.}~\bibnamefont {Dominguez}}, \bibinfo {author} {\bibfnamefont {A.~J.}\ \bibnamefont {Menssen}}, \bibinfo {author} {\bibfnamefont {D.}~\bibnamefont {Heim}}, \bibinfo {author} {\bibfnamefont {G.}~\bibnamefont {Gilbert}}, \bibinfo {author} {\bibfnamefont {D.}~\bibnamefont {Englund}}, \ and\ \bibinfo {author} {\bibfnamefont {M.}~\bibnamefont {Eichenfield}},\ }\href@noop {} {\bibfield  {journal} {\bibinfo  {journal} {Nature Photonics}\ }\textbf {\bibinfo {volume} {16}},\ \bibinfo {pages} {59} (\bibinfo {year} {2022}{\natexlab{a}})}\BibitemShut {NoStop}%
\bibitem [{\citenamefont {Dong}\ \emph {et~al.}(2022{\natexlab{b}})\citenamefont {Dong}, \citenamefont {Heim}, \citenamefont {Witte}, \citenamefont {Clark}, \citenamefont {Leenheer}, \citenamefont {Dominguez}, \citenamefont {Zimmermann}, \citenamefont {Wen}, \citenamefont {Gilbert}, \citenamefont {Englund} \emph {et~al.}}]{dong2022piezo}%
  \BibitemOpen
  \bibfield  {author} {\bibinfo {author} {\bibfnamefont {M.}~\bibnamefont {Dong}}, \bibinfo {author} {\bibfnamefont {D.}~\bibnamefont {Heim}}, \bibinfo {author} {\bibfnamefont {A.}~\bibnamefont {Witte}}, \bibinfo {author} {\bibfnamefont {G.}~\bibnamefont {Clark}}, \bibinfo {author} {\bibfnamefont {A.~J.}\ \bibnamefont {Leenheer}}, \bibinfo {author} {\bibfnamefont {D.}~\bibnamefont {Dominguez}}, \bibinfo {author} {\bibfnamefont {M.}~\bibnamefont {Zimmermann}}, \bibinfo {author} {\bibfnamefont {Y.~H.}\ \bibnamefont {Wen}}, \bibinfo {author} {\bibfnamefont {G.}~\bibnamefont {Gilbert}}, \bibinfo {author} {\bibfnamefont {D.}~\bibnamefont {Englund}},  \emph {et~al.},\ }\href@noop {} {\bibfield  {journal} {\bibinfo  {journal} {APL Photonics}\ }\textbf {\bibinfo {volume} {7}} (\bibinfo {year} {2022}{\natexlab{b}})}\BibitemShut {NoStop}%
\bibitem [{\citenamefont {West}\ \emph {et~al.}(2019)\citenamefont {West}, \citenamefont {Loh}, \citenamefont {Kharas}, \citenamefont {Sorace-Agaskar}, \citenamefont {Mehta}, \citenamefont {Sage}, \citenamefont {Chiaverini},\ and\ \citenamefont {Ram}}]{west2019low}%
  \BibitemOpen
  \bibfield  {author} {\bibinfo {author} {\bibfnamefont {G.~N.}\ \bibnamefont {West}}, \bibinfo {author} {\bibfnamefont {W.}~\bibnamefont {Loh}}, \bibinfo {author} {\bibfnamefont {D.}~\bibnamefont {Kharas}}, \bibinfo {author} {\bibfnamefont {C.}~\bibnamefont {Sorace-Agaskar}}, \bibinfo {author} {\bibfnamefont {K.~K.}\ \bibnamefont {Mehta}}, \bibinfo {author} {\bibfnamefont {J.}~\bibnamefont {Sage}}, \bibinfo {author} {\bibfnamefont {J.}~\bibnamefont {Chiaverini}}, \ and\ \bibinfo {author} {\bibfnamefont {R.~J.}\ \bibnamefont {Ram}},\ }\href@noop {} {\bibfield  {journal} {\bibinfo  {journal} {Apl Photonics}\ }\textbf {\bibinfo {volume} {4}} (\bibinfo {year} {2019})}\BibitemShut {NoStop}%
\bibitem [{\citenamefont {Castillo}\ \emph {et~al.}(2024)\citenamefont {Castillo}, \citenamefont {Shugayev}, \citenamefont {Dominguez}, \citenamefont {Gehl}, \citenamefont {Karl}, \citenamefont {Leenheer}, \citenamefont {Little}, \citenamefont {Jau},\ and\ \citenamefont {Eichenfield}}]{castillo2024cmos}%
  \BibitemOpen
  \bibfield  {author} {\bibinfo {author} {\bibfnamefont {Z.~A.}\ \bibnamefont {Castillo}}, \bibinfo {author} {\bibfnamefont {R.}~\bibnamefont {Shugayev}}, \bibinfo {author} {\bibfnamefont {D.}~\bibnamefont {Dominguez}}, \bibinfo {author} {\bibfnamefont {M.}~\bibnamefont {Gehl}}, \bibinfo {author} {\bibfnamefont {N.}~\bibnamefont {Karl}}, \bibinfo {author} {\bibfnamefont {A.}~\bibnamefont {Leenheer}}, \bibinfo {author} {\bibfnamefont {B.}~\bibnamefont {Little}}, \bibinfo {author} {\bibfnamefont {Y.-Y.}\ \bibnamefont {Jau}}, \ and\ \bibinfo {author} {\bibfnamefont {M.}~\bibnamefont {Eichenfield}},\ }\href@noop {} {\bibfield  {journal} {\bibinfo  {journal} {arXiv preprint arXiv:2407.00469}\ } (\bibinfo {year} {2024})}\BibitemShut {NoStop}%
\bibitem [{\citenamefont {Atabaki}\ \emph {et~al.}(2018)\citenamefont {Atabaki}, \citenamefont {Moazeni}, \citenamefont {Pavanello}, \citenamefont {Gevorgyan}, \citenamefont {Notaros}, \citenamefont {Alloatti}, \citenamefont {Wade}, \citenamefont {Sun}, \citenamefont {Kruger}, \citenamefont {Meng} \emph {et~al.}}]{atabaki2018integrating}%
  \BibitemOpen
  \bibfield  {author} {\bibinfo {author} {\bibfnamefont {A.~H.}\ \bibnamefont {Atabaki}}, \bibinfo {author} {\bibfnamefont {S.}~\bibnamefont {Moazeni}}, \bibinfo {author} {\bibfnamefont {F.}~\bibnamefont {Pavanello}}, \bibinfo {author} {\bibfnamefont {H.}~\bibnamefont {Gevorgyan}}, \bibinfo {author} {\bibfnamefont {J.}~\bibnamefont {Notaros}}, \bibinfo {author} {\bibfnamefont {L.}~\bibnamefont {Alloatti}}, \bibinfo {author} {\bibfnamefont {M.~T.}\ \bibnamefont {Wade}}, \bibinfo {author} {\bibfnamefont {C.}~\bibnamefont {Sun}}, \bibinfo {author} {\bibfnamefont {S.~A.}\ \bibnamefont {Kruger}}, \bibinfo {author} {\bibfnamefont {H.}~\bibnamefont {Meng}},  \emph {et~al.},\ }\href@noop {} {\bibfield  {journal} {\bibinfo  {journal} {Nature}\ }\textbf {\bibinfo {volume} {556}},\ \bibinfo {pages} {349} (\bibinfo {year} {2018})}\BibitemShut {NoStop}%
\bibitem [{\citenamefont {Baehr-Jones}\ \emph {et~al.}(2023)\citenamefont {Baehr-Jones}, \citenamefont {Ardalan}, \citenamefont {Chang}, \citenamefont {Jafarlou}, \citenamefont {Serey}, \citenamefont {Zarris}, \citenamefont {Thompson}, \citenamefont {Darbinian}, \citenamefont {West}, \citenamefont {Behnia} \emph {et~al.}}]{baehr2023monolithically}%
  \BibitemOpen
  \bibfield  {author} {\bibinfo {author} {\bibfnamefont {T.}~\bibnamefont {Baehr-Jones}}, \bibinfo {author} {\bibfnamefont {S.}~\bibnamefont {Ardalan}}, \bibinfo {author} {\bibfnamefont {M.}~\bibnamefont {Chang}}, \bibinfo {author} {\bibfnamefont {S.}~\bibnamefont {Jafarlou}}, \bibinfo {author} {\bibfnamefont {X.}~\bibnamefont {Serey}}, \bibinfo {author} {\bibfnamefont {G.}~\bibnamefont {Zarris}}, \bibinfo {author} {\bibfnamefont {G.}~\bibnamefont {Thompson}}, \bibinfo {author} {\bibfnamefont {A.}~\bibnamefont {Darbinian}}, \bibinfo {author} {\bibfnamefont {B.}~\bibnamefont {West}}, \bibinfo {author} {\bibfnamefont {B.}~\bibnamefont {Behnia}},  \emph {et~al.},\ }\href@noop {} {\bibfield  {journal} {\bibinfo  {journal} {Optics Express}\ }\textbf {\bibinfo {volume} {31}},\ \bibinfo {pages} {24926} (\bibinfo {year} {2023})}\BibitemShut {NoStop}%
\end{thebibliography}%

\end{document}